\documentclass[11pt]{article}
\pdfoutput = 1
\usepackage{jcappub}

\usepackage{amsmath,amssymb,graphicx,subfigure,pifont}
\usepackage{mathabx}
\usepackage[table]{xcolor}
\usepackage{enumitem,setspace}
\usepackage[export]{adjustbox}
\usepackage[font={small}]{caption}
\usepackage[right,modulo]{lineno}
  
\newcommand{\dx}{\mathrm d}
\newcommand{\ddx}[2]{\frac{\mathrm d #1}{\mathrm d #2}}

\newcommand{\expv}[1]{\left\langle #1 \right\rangle}

\newcommand{\matzero}{\setbox 0 \hbox{$\phantom{-1}$} \rlap{\hbox to \wd 0 {\hss0\hss}} \box 0}

\newcommand{\subtext}[2]{#1_{\text{#2}}}

\newcommand{\abs}[1]{\left\lvert #1 \right\rvert}

\newcommand*{\rom}[1]{\expandafter\@slowromancap\romannumeral #1@}
\newcommand{\RNum}[1]{\uppercase\expandafter{\romannumeral #1\relax}}

\definecolor{gold}{rgb}{1,0.843,0}
\definecolor{offwhite}{rgb}{0.85,0.85,0.85}
\definecolor{midgray}{rgb}{0.6,0.6,0.6}

\begin{document}

\title{How generic is eternal inflation?}
\author{Ross N. Greenwood}
\emailAdd{ross.greenwood@ucsc.edu}
\author{and Anthony Aguirre}
\emailAdd{anaguirre@ucsc.edu}
\affiliation{University of California Santa Cruz}

\abstract{
Everlasting inflation has far-reaching
implications for cosmology and the standing of self-localizing inferences made by observers, which have been subjects of renewed interest in light of the growing acceptance of theory landscapes and the attendant anthropic arguments.
Under what assumptions and to what extent does inflation \emph{generically} produce an eternal ``multiverse,'' without fine-tuning with respect to measures over the space of inflationary cosmologies driven by a single minimally coupled scalar field?  We address this and related questions with numerical simulations of inflationary dynamics across populations of randomly generated inflation models, instantiating a few particular simply-defined measures.}

\maketitle
\flushbottom

\setcounter{tocdepth}{0}

\section{\color{black} Introduction}
\label{sec:intro}
With the standard model of cosmology firmly established and inflation widely accepted as a core component of that model, we have come to a peculiar place. 
Many of inflation's architects hold \cite{LINDE1983177,Guth2007,Vilenkin:1983xq}
that a {\em generic} consequence of inflation is that it is \textit{eternal} or \textit{everlasting}, meaning that there is a coordinate system in which the 3-volume of the Universe increases quasi-exponentially forever, there are future-directed worldlines of infinite proper time threading inflating regions, and there are an unbounded number of thermalized post-inflationary regions ---potentially with different cosmological properties (see, e.g.\ \cite{vaas_aguirre07} for a review). 
If interminable proliferation of causally disconnected regions really were a difficult-to-avoid consequence of inflation, 
then this would make the ``multiverse"\footnote{Our use of ``multiverse'' here and henceforth refers to the Level II multiverse in Tegmark's classification scheme---causally-disconnected regions situated within the same base manifold.} the {\it de facto} standard cosmology, ushering in a host of difficulties \cite{Olum2012,Albrecht:2012zp} (and opportunities, to those so inclined \cite{Aguirre:2010rw,Nomura:2011dt,Bousso:2011up}) in the interpretation of cosmological predictions.


Any estimate of the \textit{likelihood} of eternal inflation, conditioned on observations of the Cosmic Microwave Background (CMB) and large-scale structure,
would inform the credibility of the multiverse picture and of anthropic reasoning at a cosmic scale.
{In the present work, we investigate the degree to which eternality should be considered a generic, versus finely-tuned, consequence of inflation}, with respect to various measures one might adopt over the combined space of model parameters and ``initial conditions'' for the Universe.

The paper is organized as follows:
In this section,
we review criteria for the three modes of eternal inflation, 
and discuss analytical arguments as to whether or not eternal behavior is generic.  We outline the structure of measures over single-field inflation models, and how one would term certain properties generic or fine-tuned.
In Sections~\ref{sec:methods}-\ref{sec:analysis}, we employ Monte Carlo simulations to assess the typicality of eternal inflation, computing inflationary histories across an ensemble of randomly generated potential functions, in the tradition of Tegmark's ``What does inflation really predict?'' \cite{tegmark08}.
We adopt several simple measures defined by a sampling procedure that lends itself to computational efficiency, applying statistics to the simulated data to infer rates of incidence of eternal inflation in partitions of the space of observables and scales of the potential.
In Section~\ref{sec:discussion}, we offer some concluding remarks on our findings from Monte Carlo analysis, and suggest future research opportunities.
%

\subsection{Three Roads to Eternal Inflation} \label{sec:threeroads}
We restrict attention to models in which inflation is driven by a scalar field $\varphi$ that is subject to an effective potential $V(\varphi)$ and minimally coupled to the metric, with a standard kinetic term.
%
For those models characterized by a suitable $V(\varphi)$ and initialized such that inflation can end (after at least $\sim$70 $e$-folds to be observationally viable), is the inflating physical 3-volume on space-like hypersurfaces in the future of a finite initial volume bounded in every coordinate system?  If not, then inflation is ``eternal.''


\paragraph{Stochastic Inflation}

When quantum fluctuations of the inflaton field\footnote{in conjunction with a quantum-to-classical transition yielding a stochastic distribution over Hubble-sized field configurations} dominate over its classical evolution according to the slow roll equation of motion, the end of inflation is delayed indefinitely in a non-decreasing physical 3-volume.
Over the passage of a Hubble time $H^{-1}$, the field's expectation value changes by an amount $\Delta \varphi = \lvert\dot\varphi\rvert H^{-1} = V_{,\varphi} / 3H^2 $ as it slowly rolls.  
During the same time interval, quantum fluctuations with wavelength $H^{-1}$ and amplitude drawn from a Gaussian distribution of width  $\delta\varphi_q =  H/2\pi$ may drive the field in a given Hubble volume up or down the potential slope relative to its classical trajectory.  
If in that time, the probability for the field to move higher up the slope is greater than the ratio of initial to final physical volumes ($1/e^3$), then on average at least one Hubble volume is likely to continue inflating.  
This corresponds to a stochastic eternal inflation criterion relating the effective potential and its gradient
\begin{equation}
V(\varphi)^{3/2} \gtrsim 6.6 \, \abs{V'(\varphi)} \subtext M P^3 \label{seic}
\end{equation}
Here $M_P \equiv \sqrt{\hbar c / 8\pi G}$ is the reduced Planck mass;
we generally assume natural units with $\hbar = c = G \equiv 1$.

\paragraph{Long-lived Metastable de Sitter Vacua}
\label{fvei}


In the {false-vacuum eternal inflation} scenario, the inflaton comes to occupy a local minimum of $V(\varphi)$ with a positive vacuum energy---the same initial setup as that of the Old Inflation scenario \cite{Guth:1980zm}.  As long as the field remains in that vacuum, spacetime is locally de Sitter with Hubble parameter $\subtext{H}{F} = (V(\varphi_{\text{F}})/3 M_{\text{P}}^2)^{1/2}$.
Either by tunneling through the barrier to a new field configuration and geometry, or by ascending the barrier wall by a sequence of small fluctuations, regions may ``escape'' the false vacuum to continue descent toward a lower minimum.  
%
If the transition rate is small, then the volume of space that exits the false vacuum is more than recouped by the expansion of neighboring regions that do not.  

Consider worldlines that pass through a flat hypersurface of the false-vacuum de Sitter space 
at an initial time $t_0$.  The fraction of those worldlines that pass through a locally still-inflating patch of a similarly defined hypersurface at time $t > t_0$ is
\begin{equation} f_{\text{inf}} = \exp \left[ -\frac{4\pi}{3} \frac{\lambda(t-t_0)}{H_{\text{F}}^3} \right] 
\label{eq:inf_frac} \end{equation}
where $\lambda$ is the nucleation rate per 4-volume.  The physical volume of inflating space goes like $ v_{\text{inf}} \, \propto \, f_{\text{inf}} \, e^{3H_{\text{F}}t} $, so a transition rate
\begin{equation} \lambda < 9 H_{\text{F}}^4/4 \pi \label{fv_rate} \end{equation} ensures that the 3-volume of inflating space never decreases in the 4-volume coincident with a statistically large population of initial Hubble volumes \cite{vaas_aguirre07}.  
If a transition is followed by enough slow roll inflation to hide any otherwise observable relics of bubble nucleation and solve the horizon problem, then one may have occurred in the past of our observable universe.

The case of tunneling through the barrier is treated by the Coleman-de Luccia (CDL) instanton formalism \cite{Coleman:1980aw}.
The case of stochastic fluctuation of Hubble volumes up the barrier wall (or equivalently, thermal fluctuation with a characteristic de Sitter temperature) was first explored by Hawking and Moss \cite{hawkingmoss} as the limiting case of a CDL instanton in which both termini are identified with the top of the barrier.

\paragraph{Inflating Topological Defects}

A third mechanism for eternal inflation operates even in a classical setting: what if conditions for inflation \textit{cannot} end everywhere, due to topological considerations?
Take for example a potential with two vacua, separated by a local maximum at $\varphi = 0$
\begin{equation} V(\varphi) = \frac{\kappa}{4} \left( \varphi^2 - \frac{m^2}\kappa \right)^2 \label{toppotential} \end{equation}
If in separated regions of space the field has settled into different vacua at $\varphi = \pm m/\sqrt{\kappa}$, then the field must obtain the local maximum somewhere in between, forming a {\it domain wall} with a positive energy density.  
%
%
When the characteristic width of domain walls is comparable to the Hubble scale, gravitational effects can lead to {\it topological eternal inflation}.

Suppose a nearly static domain wall (necessarily of a thickness much less than the Hubble radius associated with the potential energy at the peak) is a solution of a given potential.
Given an initial Hubble volume not containing a domain wall, but in which $\varphi$ is nearly homogeneous around the top of the barrier, it is of interest under what conditions the domain wall that forms as a consequence of the field's semi-classical descent from the peak is nearly static and sub-Hubble in scale.\footnote{The alternative is that inflation of what will become the domain wall core outpaces its collapse in physical coordinates, and a localized solution is precluded by the ensuing expansion.}

A necessary but insufficient condition for a sub-Hubble defect to form is that the gradient of the field configuration around the top of a potential barrier is initially increasing in physical coordinates.
Assume that $\varphi(t,x)$ is initially linear in its spatial dependence in a small region around which it obtains the peak value, with a small proportionality factor $k(t)$, defined with respect to the physical distance $x e^{H_0 t}$: 
\[ \varphi(t,x) \approx k(t) a(t) x H_0 \]
where $H_0$ is the Hubble parameter at the peak (a convenient mass scale), $x$ is a comoving coordinate, and
$ a(t) \approx e^{H_0 t}$.
In the vicinity of the maximum we take the potential to be approximately quadratic:
$ V(\varphi) \approx V_0 - \tfrac12 \mu^2 \varphi^2 $.
The equation of motion for $k(t)$ is then
\[ \ddot k + 5 H \dot k + (4 H^2 + \dot H - \mu^2) k = 0 \]
If the expression in parentheses is positive, then $k(t)$ behaves like an overdamped harmonic oscillator, and the domain wall grows ($k(t)$ vanishes at late times).
So if $V(\varphi)$ is sufficiently flat near the maximum
\begin{equation} \frac1{\kappa V_0} \abs{ \ddx{^2 V}{\varphi^2} }_{\varphi=0} \equiv \eta_V< \frac{4-\epsilon}3 \label{eq:topcrit} \end{equation}
where $\epsilon$ is the first slow roll parameter, then the domain wall grows even if a nearly static sub-Hubble-scale domain wall solution were possible---resulting in topological eternal inflation.
Otherwise, $k(t)$ grows monotonically within the validity of this approximation, leading potentially to a collapse.\footnote{A full collapse can be thwarted by Hubble friction lower on the potential.}
%
\footnote{Even if $\varphi$ does not initially interpolate between the two basins of attraction, if it inflates from its initial configuration, then stochastic fluctuations in the population of Hubble volumes produced near the peak may be enough to drive the field in some regions over the hilltop---creating a topological defect.}

\label{sec:howgeneric}

\subsection{\color{black} The Case for \emph{Generic}}
\label{sec:caseforgeneric}

It is widely believed that eternality 
is a generic prediction of inflation---a consequence arising without the need for fine-tuning of model parameters or initial conditions.
It is our impression that the prevailing rationale for that belief is largely qualitative, and based on a heuristic sampling of inflation models with substantial coverage in the literature. 
Nonetheless it appears well motivated; what---in rough sketch---is the basis for it?

%
By inspection of Eq.~\eqref{seic}, stochastic fluctuations can compete with slow roll in models in which $\varphi$ is initialized near an inflating local maximum or saddle point and for those in which $V(\varphi)$ is large in the domain of initial field values (e.g. inflation in a quadratic potential with mass parameter $m$ and field excursion $\Delta\phi \gtrsim 4 (M_{P}^3/m)^{1/2}$).
Slow roll already requires a very flat potential; stochastic eternal inflation needs only somewhat flatter or more energetic sites of inflation. 
Plausible extensions to the standard treatment of fluctuations---like warm inflation \cite{Berera:1995ie} with its account of thermal effects---increase the amplitude of fluctuations of the inflaton, shrinking the gap between sufficient conditions for successful inflation and eternal behavior.
If we are assumed to occupy a region of model space in which inflation is generic, it seems to follow from the above considerations that eternal inflation should not be much less difficult to avoid.
The topological inflation criterion given by Eq.~\eqref{eq:topcrit} is always satisfied for a single inflaton initialized at a maximum where the second potential slow roll condition $\eta_V \ll 1$ is met.  The latter condition---though not required for some quasi-exponential expansion to occur---is almost always needed to solve the horizon problem\footnote{This is relevant even when initializing at a maximum, as quantum or thermal fluctuations render finite the expected time and $e$-folds elapsed in the vicinity of the peak.}.
Quantum fluctuations dominate the slow evolution of $\expv\varphi$ near the peak, resulting in an inhomogeneous field configuration with $\expv\varphi$ descending toward the minima of both conjoining half-basins, separated in space by an inflating topological defect that can never be excised.
{Even if $\varphi$ is initialized only near a maximum, stochastic inflation always occurs within a neighborhood of the potential peak, and fluctuations can drive some regions over the peak to descend toward the adjacent minimum, producing a defect.  This can occur by chance even if the formal stochastic inflation criterion is not met at the initial field value.}

If we take the setting of inflation to be a ``landscape'' potential and populate the presumably large number of false vacua therein, then we get eternal inflation almost no matter what.
%
If the barrier between false and true vacua is broad $(\abs{V''} \ll V/m_{\text{P}}^2)$, then the largest contribution to the transition comes from the Hawking-Moss (HM) instanton.
We adopt the interpretation of the HM calculation \cite{hawkingmoss} as yielding the rate at which Hubble volumes occupying the false vacuum basin thermally fluctuate into the true vacuum basin, with energy comparable to the height of the barrier \cite{A07transitionsbetween}.
Since the HM transition with a slow rate of diffusion invokes stochastic inflation---the field undergoes a series of quantum or thermal fluctuations up the barrier wall---it follows that all models with a slow HM transition are necessarily eternal.  


\subsection{The Dissent}
\label{sec:dissent}

Two schools of criticism of eternal inflation are represented in the literature.
One attacks the foundational assumptions on which the established treatments outlined in \S~\ref{sec:threeroads} are based;  
for review of several such arguments, see e.g.\ \cite{Mukhanov:1996ak,mersiniparker07,Parker:2007ni,Nelson:2016kjm,Boddy:2016zkn}.
Another accepts the established treatments but questions the predominance of the associated criteria in the space of inflation models consistent with observation,
honing in on several hints of fine-tuning.

The primordial perturbation spectrum inferred from measurements of the CMB implies a significant scale discrepancy between a regime of stochastic inflation, in which curvature perturbations $\delta\mathcal R/\mathcal R$ are of order unity, and horizon exit of modes of observational relevance.
Suppose we parameterize the inflaton potential in terms of the vertical and horizontal mass coefficients $m_v$ and $m_h$, along with a dimensionless function $f(x)$ of order unity defining its shape:
\[ V(\varphi) \equiv (m_v m_{\text{P}})^4 f \!\left({\varphi}/(m_h m_{\text{P}}) \right) \]
Here $m_{\text{P}}$ is the Planck mass.
With the measured magnitude of scalar perturbations $Q_s \sim \mathcal O(10^{-5})$, the quantity $f(x)^{3/2} f'(x)^{-1}$, appearing in both the expression for $Q_s$ and Eq~\eqref{seic}, must shrink by a factor $\mathcal O(10^{-3})$ between a scale of stochastic eternal inflation and the horizon exit scale.
In this sense the sufficient conditions for inflation consistent with the data do not strongly constrain the part of the potential relevant for stochastic eternal inflation.
%
%
In \cite{kinneyfreese15}, this scale discrepancy is cast as an upper bound $\alpha_{\text{max}}<0$ on the running of the spectral index, below which inflation is non-eternal (assuming higher order ``runnings-of-the-running'' can be neglected).
\begin{equation} \alpha < \frac{(1-n_s)^2}{4\ln \mathcal P_{\mathcal R,\star}} \approx -4 \times 10^{-5} \label{runningbound} \end{equation}
%
If the field history does not traverse intervals on the potential with vastly different characteristics before the end of inflation, then there are large portions of parameter space consistent with observation that do not produce stochastic eternal inflation.

If the barrier separating true and false vacua turns over sharply (large $V''/V$), then the largest contribution to a vacuum transition is a Coleman-de Luccia instanton, which tends to terminate very close to the true minimum with a thin-walled bubble.  While instanton transition rates are generically much smaller than the upper bound in \eqref{fv_rate}, one would need an instanton that terminates atypically high on the slope of a nearly-Minkowski half-basin, separated from the minimum by an interval of field space in which the potential is remarkably flat by the standards of the potential barrier. 
However, the rarity of coincidence of such features of the potential may be balanced by conditioning on inflation producing enough $e$-folds to begin with.  
Even if the inflaton is initialized in a false vacuum, our prior expectation on the number of $e$-folds is already high to be consistent with observation; that successful inflation following a CDL transition is rare is for all intents counterfactual.

Before conditioning on a small final vacuum energy, one might suspect models with successful inflation following a Hawking-Moss transition of being unnatural, as the potential featuring a broad barrier must also vary quickly before the minimum in order to give a clean exit from inflation. 
This is alleviated only somewhat if we fix the energy of the true vacuum to be very small, so that inflation always ends close to the minimum where $\epsilon_V \sim 1$.
This puts a Hawking-Moss transition among field histories yielding sufficient inflation with a viable scalar amplitude and spectral index, but a typically too large tensor-to-scalar ratio.

\subsection{Measures on Cosmologies}
\label{sec:measures}

What would it {\em mean} for eternal inflation to be generic?  As a matter of history, it has meant that of the inflation models that have been devised, {many} appear to be eternal in regions of the space of parameters and initial conditions that have warranted scrutiny.  More properly, it should mean that {given} some representative {\it measure} (or class of measures) over inflation models and initial conditions, those combinations resulting in eternal inflation comprise a large fraction of the measure, or perhaps of the measure over combinations that lead to observationally viable cosmologies. 
%
%

One might argue that even if eternal inflation does not dominate a measure over model space, those models with eternal inflation produce vastly more variety in thermalized regions, and so an observer is warranted in assuming that the conditions for eternal inflation are in their region's past.
This stance is sensible only if one grants that a population of models\footnote{or of initial conditions on a vast potential with both eternal and non-eternal inflation} is in fact realized, so that they are in effect competing dynamically for representation in a final ensemble.  
%
For this study, we rather assume that one potential is actually realized, and probe the likelihood that a Hubble volume with initial conditions sampled from a particular distribution undergoes eternal inflation.

{\it Genericity} might refer to a posterior distribution over a collection of parameters $\mathbf p_{\text{eternal}}$ characterizing eternal inflation, given priors on those parameters, observational data $\mathbf d$, and a dictionary of correlations between the parameters modeling the data and the necessarily hidden parameters in $\mathbf p_{\text{eternal}}$ (we only have access to our one observable universe).  Let the full set of parameters be denoted by
$\mathbf p = \mathbf p_{\text{eternal}} \bigcup \mathbf p_{\text{obs}} $,
where $\mathbf p_{\text{obs}}$ includes those parameters modeling observables from any successful inflation model that are accessible to our instruments.\footnote{See \S~\ref{sec:simflow} for definitions of the quantities making up $\mathbf p_{\text{obs}}$.} 
Since our model of the data can only connect $\mathbf d$ to $\mathbf p_{\text{obs}}$, the posterior distribution over the hidden parameters in $\mathbf p_{\text{eternal}}$ is determined entirely by how they correlate with those in $\mathbf p_{\text{obs}}$.
%
For each measure $m$, there is a distribution $f_m(\mathbf p \mid \mathbf p_{\text{obs}})$ connecting the hidden eternal sector to parameters that make contact with the data.
The probability associated with the full parameter vector $\mathbf p$ is then
\begin{equation} p_m(\mathbf p \mid \mathbf d) = \int f_m(\mathbf p \mid \mathbf p_{\text{obs}}) \,\ell(\mathbf p_{\text{obs}} \mid \mathbf d) \, \dx \mathbf p_{\text{obs}}. \label{likelihood} \end{equation}
Generic eternal inflation would mean that $p_m(\mathbf p \mid \mathbf d)$ nearly vanishes outside of the region of parameter space spanned by $\subtext{\mathbf p}{eternal}$ that is labeled eternal.
%

From a frequentist point-of-view, one is concerned with estimating rates of incidence of quantities characterizing eternal inflation meeting predefined thresholds (see criteria in \S~\ref{sec:threeroads}), rather than with the full posterior distributions over those quantities.
Considered in these terms, \textit{generic} could also mean that an estimate of the rate or probability of occurance of eternal inflation is close to 1.
This definition has the advantage of not requiring one to presume a model for the distributions of parameters in $\mathbf p_{\text{eternal}}$, but rather only that of a rate parameter which can be taken to be beta-distributed.
Providing the means to compute these rates in a crude form is in large part the aim of this study.

\section{Monte Carlo Methods}
\label{sec:methods}


Supposing a scalar field driving inflation was governed by an effective potential of whose origin we are ignorant, it would be worthwhile to discover with what probability (and observational correlate) we encounter eternal behavior, subject to measures admitting a high degree of variability in the potential. 
%
With a well behaved distribution, we could make some headway with a purely analytical approach, extending the results of works like \cite{Rudelius2019};
but having a trove of simulated data affords a freedom to make arbitrary cuts on observables.
Since we are most interested in models that come close to producing observationally viable cosmologies and require the means to select for them,
we find a numerical approach to be of greater marginal value for this study.  

\subsection{Desperate Measures}
\label{sec:desperatemeasures}


If there is a \textit{true} measure over inflationary cosmologies, it is of course unknown; nonetheless, we argue that 
even simple measure prescriptions can offer insight into what might be deemed generic versus fine-tuned.
From a stance of humility in light of the issues discussed at length in \cite{tegmark08} facing all would-be measure bearers, we aim to travel a middle road of devising measures that are easily computable and provide adequate coverage of model-space, while not trying too hard to justify a particular measure on specific physical grounds.

\subsubsection{\color{black} Sampling Potential Functions}

Following the example of Tegmark \cite{tegmark08}, we draw effective potential functions as one-dimensional Gaussian random fields (GRFs). 
The GRF has several properties that make it suitable for this role: it is smooth and continuous, bounded from above and below, and its statistics are translation invariant.\footnote{The last stands in contrast to the one-parameter space of quadratic potentials, for example, which is guaranteed to have special behavior near $\varphi = 0$.}
%
(Furthermore, many recent works employ a GRF scalar field potential as a placeholder for the distribution output by a string theory landscape.)
We express the potential $V(\varphi)$ in terms of a dimensionless GRF
\begin{equation} f(x) = \frac{a_0}{\sqrt 2} + \sum_{k=1}^{\subtext{k}{max}} a_k \cos(kx) + \sum_{k=1}^{\subtext{k}{max}} a_{-k} \sin(kx) \label{grf} \end{equation}
and constants defining the mass scales of the potential ($m_v$) and the inflaton ($m_h$) in the potential.
\begin{equation} V(\varphi) = (m_v m_{\text{P}})^4 f ( (m_h m_{\text{P}})^{-1} \varphi ) \label{eq:grfV} \end{equation}
Each coefficient $a_k$ is sampled from a Gaussian distribution such that 
\begin{equation} \text{Var} (a_k) = q^\gamma e^{-q^2/2}, \qquad q \equiv {k}/{\sqrt{k_{\text{max}}}} \label{grf_cov_var} \end{equation}
Tegmark found that varying the scale dependence of $a_k$ through the shape parameter $\gamma$ did not produce interesting discrepancies in the resulting distributions, so we take $\gamma = 0$ for our analyses unless otherwise specified.
We generally take $k_{\text{max}} = 30$.

The function $f(x)$ defines the shape of the potential, but we must also impose priors on the vertical and horizontal mass scales---$m_v$ and $m_h$, respectively---to enact the full measure, and to compute statistics of subsamples aggregated from multiple mass pairings.
For most results, we assume a prior for each that is uniform on a log-scale within some designated mass ranges; but we also consider a straight uniform prior, which gives greater weight to larger mass scales for the field that tend to result in more $e$-folds.
Beyond this base prior, we adopt two schemes for constructing measures from grids of mass scales---\emph{democratic} and \emph{epektacratic} (rule by expansion)---elaborated in Appendix \ref{app:stats_methods}.


We would ultimately condition on the smallness of the vacuum energy at the stable minimum where inflation ends.  Rather than sampling from the full distribution and then conditioning on $\rho_\Lambda$ being many orders of magnitude smaller than the scale of the potential, we aim for a shortcut to approximate such a move without covering the vast regions of parameter space in which the vacuum energy is negative or significantly too large.
Very simply, we first check whether any minimum in the search space has a vacuum energy within $\pm0.01 \, m_v^4 m_{\text{P}}^4$; if so, then we shift the whole potential to bring the vacuum energy in that basin to zero.
Inflation must end in that particular basin in order for that model to be counted. 
%



\subsubsection{Sampling Initial Conditions}
\label{sec:sampleinits}
We are concerned with the 4-volume in the future of one initial Hubble volume;
however, a useful measure must entail assumptions about nearest neighbor Hubble volumes. We assume nearly homogeneous initial conditions at the Hubble scale.
Over initial values of the homogeneous inflaton field $\varphi$,
Tegmark adopted two measures:
\begin{description}[leftmargin=0.55cm]

\item[A] { Sample field values maximizing $V(\varphi)$, weighted by the distance in field space between the two adjacent minima.  (This is equivalent to sampling uniformly and then going uphill to the peak.)  Discard instances in which $\abs{\eta_V} > 1$ at $\varphi_0$.} 

\item[B] { Sample field values uniformly.  
Discard instances in which $\epsilon_V > 1$ or $\abs{\eta_V} > 1$ at $\varphi_0$.
}
\end{description}
We adopt these two measures and add a third:
\begin{description}[leftmargin=0.55cm]

\item[C] { Sample field values a distance in field space equal to $H_{\text{max}}/2\pi = \sqrt{2 V(\phi_{\text{max}})/3\pi}$ from local maxima of $V(\varphi)$, weighted by the distance in field space between the two adjacent local minima.  Discard instances in which $\epsilon_V > 1$ or $\abs{\eta_V} > 1$ at $\varphi_0$. }

\end{description}
Measure C is equivalent to sampling from Measure A and then adding a standard deviation of the slow roll stochastic fluctuation of Hubble-scale modes to that sampled value.  We consider it as a possibly interesting interpolation between Measures A and B that excludes the field space interval in the small neighorhood of the peak in which inflation is nearly a given.


All three measures as framed above require that both slow roll conditions are satisfied at the starting point, and then take \begin{equation} \ddot\varphi = 0 \qquad \dot\varphi = -M_{\text{P}} V'(\varphi)/\sqrt{3 V(\varphi)} \label{sr_profile} \end{equation}  
More properly, we would sample from a well-motivated distribution over $\dot\varphi$ and higher derivatives, and integrate the full equations of motion; $\varphi$ could then barrel through a short enough interval where the potential slow roll conditions are met without inflation taking place.  
As Tegmark pointed out, if 
\begin{equation}
\dot\varphi \lesssim \sqrt{2V(\varphi)}
\label{slowrollattractor}
\end{equation}
in an interval where $\epsilon_V,\abs{\eta_V} < 1$, then the full equation of motion exhibits an attractor behavior leading quickly to the slow roll profile \eqref{sr_profile} \cite{liddle2000cosmological}.
Sampling $\dot\varphi$ from a distribution and then conditioning on slow roll, the population of surviving models would be those that approximate the above measures, with additional weighting like $m_v^2/\langle{\dot\varphi^2}\rangle^{1/2}$.
Since we are drawing from an array of mass scales, we could model such effects from our simulated data without running the full dynamics, by adjusting the prior on $m_v$.  (Our results presented below do not take into account such a reweighting.)

\paragraph{Inflation Below the Peak}
Stochastic inflation occurs in every model sampled to reflect Measure A that produces observables, since there is always an inflating interval contiguous with the maximum where $V'(\varphi)$ vanishes.  
If we waive the requirement that inflation continues through 70 $e$-folds \emph{from the maximum}, and instead let inflation start lower on the potential if the potential slow roll conditions are met with $\varphi$ varying slowly enough at the start of a slow roll interval, we allow for the possibility of non-stochastically eternal inflation initialized at a non-inflating peak.
We only require that $\dot\varphi$ is below the bound of the slow roll attractor when it reaches the top of an interval in which $\epsilon_{V},\abs{\eta_V} < 1$.
With these considerations, we adopt the following modified version of Measure A:
\begin{description}[leftmargin=0.75cm]

\item[A$^*$] Sample field values maximizing $V(\varphi)$, weighted by the distance in field space between the two adjacent minima.  If  $\eta_V > 1$ at the peak, then assume inflation starts where $\epsilon_{V},\abs{\eta_V} < 1$ first becomes valid $(\varphi_{\text{sr}})$, if $\dot\varphi_{\text{sr}} < \sqrt{2V(\varphi_{\text{sr}}})$ along a trajectory approaching the peak as $t \to -\infty$. (For details of the calculation, see Appendix~\ref{app:mc_methods2}.) 


\end{description}

\subsection{Simulation Design}
\label{sec:simflow}

Our design intent is to characterize the evolution of the inflaton---including up to 1 Coleman-de Luccia or Hawking-Moss transition event---accurately enough to inform the distributions $f_m(\mathbf p \mid \mathbf p_{\text{obs}})$ from Eq.~\eqref{likelihood}, while exploiting justifiable shortcuts in order to economize on computing time.
For each instance toward building up the distribution, the steps are outlined in Appendix~\ref{app:sim_design}.
We record observables\footnote{$Q_s$ and $n_s$ are the scalar amplitude and spectral index, respectively; $\alpha$ is the running of $n_s$; $r$ is the tensor-to-scalar ratio; $n_t$ is the tensor spectral index; $\Omega_{\text{tot}}$ is the critical density fraction; $\rho_\Lambda$ is the vacuum energy in the potential basin where inflation ends.} if inflation ends with $\mathcal N_e > 70$ in a vacuum with $\rho_\Lambda \ll 1$, along with indicators for eternal inflation:
\begin{align} 
\mathbf p_{\text{obs}} &= (Q_s,r,n_s,\alpha,n_t,\delta\rho/\rho,\log \abs{\Omega-1}) \label{eq:paramvec} \\
\mathbf p_{\text{eternal}} &= (N_s,\langle\mathcal  N_{e,\text{stoch}}\rangle,b_t,\lambda_{\text{fv}},H_{\text{F}},b_{\text{HM}}) \label{eq:paramvec2}
 \end{align}
%
%
In \eqref{eq:paramvec2}, $N_{\mathrm s}$ is the number of dis-contiguous field space intervals in which the stochastic eternal inflation criterion is valid for at least one slow-roll $e$-fold\footnote{A very small field space interval in which \eqref{seic} is satisfied---but in which a slowly rolling inflaton would not drive even a single inflationary $e$-fold as it traversed the interval---is not counted as producing stochastic eternal inflation.};
$\langle\mathcal  N_{e,\text{stoch}}\rangle$ is the sum of ratios of the widths of stochastic inflating intervals in field space to the amplitudes of quantum fluctuations characteristic to those intervals\footnote{This gives a measure of how easily the field can fluctuate out of the conditions for stochastic inflation, in the case of $N_e = 1$.  This metric is not used in our analysis, but is included in the collected data sets.}; 
$b_{\mathrm t}$ is a Boolean flag for topological eternal inflation;
$\lambda_{\text{fv}}$ is the rate of quantum diffusion from a metastable false vacuum; 
$H_{\text{F}}$ is the Hubble parameter in that false vacuum;
and $b_{\text{HM}}$ is a Boolean flag indicating whether the transition is dominated by the Hawking-Moss instanton.

In terms of the potential $V(\varphi)
$; the slow roll parameters $\epsilon_V \equiv (V'/V)^2/2$, $\eta_V \equiv V''/V$, and $\xi_V \equiv V'''/V$; and the number of $e$-folds before horizon exit $\mathcal N_{e,\text{before}}$: the quantities in $\mathbf p_{\text{obs}}$ are given by
\begin{align*}
Q_s^2 &= V(\subtext\varphi{exit}) /(150\pi^2 \, \epsilon_V \,\kappa \subtext{m}{P}^2) \\
r &= 16 \epsilon_V \\
n_s &= 1-6\epsilon_V+2\eta_V \\
\alpha &= 16 \epsilon_V \eta_V - 24 \epsilon_V^2 - 2 \xi_V^2 \\
n_t &= -2 \epsilon_V \\
\delta\rho/\rho &= \log_{10} (V(\varphi_{\text{exit}})/V(\subtext{\varphi}{end})) \\
\ln {\abs{\Omega-1}} &= \ln (V(\subtext\varphi{start})/V(\subtext\varphi{exit})) - 2\mathcal N_{e,\text{before}}
\end{align*}

\section{Results \& Discussion}
\label{sec:analysis}


After binning models in the space of measure parameters and/or observables, we take the number of eternal models observed in each bin to be a binomial-distributed random variable, with a deterministic but unknown probability of eternality $\lambda$ for each bin.
Our task is then to estimate the rate $\lambda$ characterizing the bin population.
%
The measure parameters 
are the masses identifying the scales of the field ($\varphi \sim m_h$) and the energy density ($V \sim m_v^4$), along with the shape parameter $\gamma$.  
We sample $m_h$ from within a few orders of magnitude of the Planck mass, which of course limits the scope of applicability of our results to a small subspace of conceivable models.
This choice was informed by noting for which field scales we are likely to get a large enough subpopulation of inflation models that are observationally viable (see Figure~\ref{fig:142_Q}).
Taking the potential to vary on field scales within the range $\mathcal O(10^{-2}) \lesssim m_{\text{h}} \lesssim \mathcal O(10)$, we sample $m_v$ from a range in which the amplitude of scalar perturbations $Q_s$ is most likely to be consistent with the \emph{Planck 2018} data set: $\mathcal O(10^{-5}) < m_v < \mathcal O(10^{-2})$.





\subsection{Measure A: Summits}
\label{sec:MeasureA}
If the second potential slow roll condition is met at the peak, then inflation always thwarts the collapse of any initially near-homogeneous field configuration into a quasi-static domain wall, and so continues in perpetuity.
In those models, inflation is eternal by both stochastic and topological modes: quantum fluctuations dominate near the peak where $V'(\varphi)$ vanishes, and causally disconnected regions descending toward different minima of the potential are separated by an inflating domain wall.

If $M_P^2 \,V''/ V \lesssim -4/3$ in a suitably large interval around the top of the potential barrier, then small inhomogeneities around the peak tend to grow with time in physical coordinates (see \S~\ref{sec:threeroads}).
The field's potential energy will not tend to dominate its kinetic energy for a sustained bout of inflation, and
slow roll does not persist at the peak. But the model
has a chance to accrue many $e$-folds lower on the slope of the Minkowski half-basin---if the field velocity is small enough in a field space interval in which $\epsilon_V,\abs{\eta_V} < 1$---and go on to produce a viable cosmology.  
Absent these latter conditions, and if the scale of the domain wall around the sharp peak is sub-Hubble, 
no thermalized regions have enough $e$-folds of inflation in their past to solve the horizon problem.

Let us identify nested subsets of models belonging to a sample population from Measure A:
\begin{itemize}
\item Let $A$ denote the set of all models in the sample from Measure A.
\item Let $S \subset A$ denote the set of models that have {\it successful} inflation, meaning greater than 70 $e$-folds accrued in an interval in which the potential slow roll conditions are satisfied.
\item Let $D \subset S$ denote the set of models that are successful AND in which the only sustained bout of inflation occurs in a field space interval that is not contiguous with the peak.
\item Let $D' \subset D$ denote the set of models in $D$ for which the stochastic inflation criteria are never satisfied.
(All models in $S$ but not in $D'$ are stochastically eternal.)
\end{itemize}


%

\subsubsection{Stochastic Eternality}
Models in sample subset $D$ are those in which both potential slow roll conditions are met only below the potential maximum, with enough inflation in that interval to solve the horizon problem.
Only if Eq.~\eqref{seic} is satisfied on the slow roll interval lower on the potential does stochastic eternal inflation occur; if not, then the model is also in $D'$.
It is typical for classical trajectories initialized a 1-$\sigma$ fluctuation away from the peak with zero field velocity to undergo slow roll for several $e$-folds along the descent, despite the potential slow roll conditions not being met.  
Stochastic eternal inflation near the maximum is avoided when fluctuations add up around the peak to produce inhomogeneities, which are amplified by the large second derivative of the potential. 
Suppose this can result in a terminal, short-lived bout of inflation within the interval of field space around a sharply curved potential peak.

\begin{figure}[t]
\centering
\includegraphics[width=12.5cm]{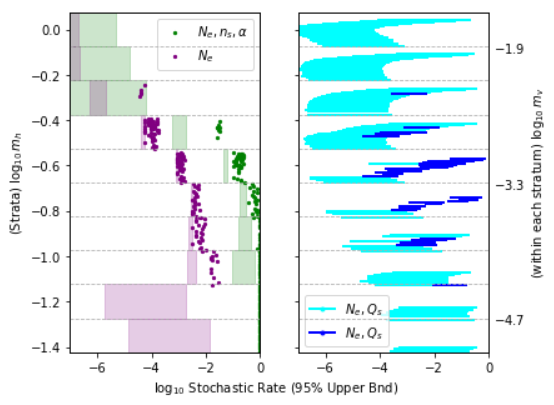} 

\caption[Stochastic inflation in Measure A]{
Incidence rate of models with slow roll starting below the peak with no stochastic inflation, among models in Measure A with 70+ $e$-folds.
Each data point represents one batch of simulations with particular $m_v,m_{\text{h}}$ (only showing batches with at least one positive event per sample).  All models within a vertical stratum have the same value of $m_h$ (the center of the stratum on the left axis); vertical position within the stratum reflects $\log_{10} m_v$ (range shown on the right axis).
Data points reflect 95\% confidence upper bounds.
In the left plot, green data points are derived from samples conditioned on $n_s$ and $\alpha$; purple data points are conditioned only on minimal $e$-folds; the shaded bars indicate 90\% confidence intervals taking models from all values of $m_v$ as belonging to one sample. 
In the right plot, the blue lines show 90\% confidence intervals derived from samples of successful models conditioned further on $Q_s$.  Darker lines reflect samples that have at least one non-stochastic model in the sample, whereas lighter points are determined only by sample size.}
\label{fig:scatter_144_145}
\end{figure}

%
In Figure \ref{fig:scatter_144_145}, we depict rates of incidence of models with $\eta_V < -4/3$ at the initial peak, in which successful slow roll inflation could begin lower on the potential without meeting the stochastic inflation criterion. 
Considered as a frequentist ratio, the numerator and denominator for each data  point in Figure~\ref{fig:scatter_144_145} are the sizes of $D' \cap \{m_v,m_h\}$ and $S \cap \{m_v,m_h\}$, respectively, where $\{m_v,m_h\} \in A$ is the subpopulation simulated with a particular pairing of mass scales.
%
Stochastic eternal inflation is generic at large field scales $m_h > 1$, where $\eta_V \sim (m_{\text{h}} m_{\text P})^{-2}$ is easily within the bound of the slow roll approximation at the peak.
This continues to low rates $\abs{D'}/\abs{S}$ in Figure~\ref{fig:scatter_144_145} for large $m_{\text{h}}$.  

The quantity of our ultimate interest is the probability of an observationally viable model undergoing inflation that is not stochastically eternal at or below the local maximum:
\[ P(m \in D' \mid m \in S \cap \{Q,n_s,\alpha,r,n_t,\dots\} ) \]
where the latter set $\{Q,n_s,\dots\}$ contains models that satisfy constraints on observables.
The probabilities represented by the purple (dark gray) data points in the left plot of Figure~\ref{fig:scatter_144_145} are
\[ P(m \in D' \mid m \in S \cap \{m_v, m_h\}) \]
Below the Planck scale $m_{\text{h}} \sim 1$, incidence of non-stochastically-eternal models among those with enough $e$-folds goes roughly as a power law with the scale of the inflaton, before conditioning on spectral features.
The data points indicate 95\% confidence upper bounds for those mass bins in which at least one non-stochastic model was observed.

\paragraph{Conditioning on Spectral Shape}
What is the effect of requiring that the angular scale dependence of the scalar CMB spectrum is consistent with the {\it Planck} 2018 data set, resulting in a scalar index $n_s$ and running $\alpha$ within their respective 95\% confidence intervals?\footnote{The running $\alpha$ is generically within the bounds from measured data after conditioning on $n_s$.}
At all field scales shown in Figure~\ref{fig:scatter_144_145}, the upper bound on rates of non-stochastically eternal inflation is greater by 1-2 orders of magnitude after conditioning on $n_s$ and $\alpha$; these figures represent the conditional probabilities:
\[ P(m \in D' \mid m \in S \cap \{m_v,m_h\} \cap \{n_s,\alpha\} ) \]

Aggregating all vertical mass scales for a given field scale, we report the resulting 90\% confidence intervals represented by the green (light gray) shaded bars in the left of Figure~\ref{fig:scatter_144_145}.
(We also show the same results for models conditioned only on number of $e$-folds (purple / light gray).)
From this we infer that non-stochastically-eternal inflation falls short of being generic with high likelihood down to below $m_h = 0.1$, among models conditioned on $n_s$ and $\alpha$.  
For smaller field scales in this regime, what we have called non-stochastic-eternal inflation (which includes some generous assumptions that cannot be taken for granted) is neither generically present nor absent.

\paragraph{Conditioning on Spectral Amplitudes}
Turning attention to the righthand plot in Figure~\ref{fig:scatter_144_145}, we find that stochastic inflation is less prevalent at larger potential scales among models with viable spectral amplitudes, in comparison to all models in $S$.  
A larger scale for the potential correlates with a greater proportion of models with inflation below the peak, among successful models satisfying constraints on the scalar amplitude $Q_s$.
This may be somewhat surprising considering that the stochastic inflation criterion is a lower bound on the amplitude of scalar curvature perturbations, which scales with $m_v^4$.
But it makes sense when considered as an effect of selecting for a suitably small scalar amplitude born of a delayed inflationary epoch.

A larger Hubble parameter $\sim m_v^2$ means greater friction, allowing the potential to have a large curvature at the peak while tending toward smaller separation in field space between the peak and an interval of slow roll (greater allowance for $V'(\varphi)$), as well as a smaller field velocity at the top of that interval.
When the potential is large, the requirement of a small amplitude for scalar perturbations means inflation must end low on the potential slope, which is more likely in models with inflation starting below the peak.
With greater potential energy the upper bound on $\dot\varphi$ for the slow roll attractor is greater, allowing more models to accrue many $e$-folds below the peak.

\subsubsection{Topological Eternality}
%
For models in the sample subset $D$, so long as $\varphi$ somewhere obtains the peak value, there is a Hubble-sized region at the top of the slow roll interval below the peak that must pass through an inflationary epoch, as the field therein descends the slope toward the Minkowski vacuum.
This is the case in which the characteristic scale of domain wall thickness is greater than the characteristic Hubble scale in the wall's core;
even as small inhomogeneities around the peak are initially magnified, the wall is supported against collapse to sub-Hubble scales by the shape of the potential, and we still end up with inflating defects interpolating between the two vacua.
Does this imply that all successful delayed inflation in Measure A
is always topologically eternal, since the intervening space separating regions occupying the vacua is inflating?  Almost (but not quite) certainly.

Suppose that in Regions I, II, and III, the field occupies the Minkowski half-basin, the neighborhood of the sharp barrier, and the adjacent basin, respectively.
Only if Region III grows to a size larger than the Hubble scale in that basin---effectively ``pinning'' the de Sitter horizon surrounding that region to a value in the adjacent basin---do we get a {\it topological} defect.
So long as the defect is contained within a single de Sitter horizon, one can entertain the possibility of a nonperturbative fluctuation that could put the field in Regions I-III in the true-vacuum basin, and make an end to inflation possible.

\subsection{Measure B: Uniform}
Measure B draws initial field values uniformly; 
because the statistical behavior of GRFs is translation-invariant, this is equivalent to simply choosing $\varphi_0 = 0$ for each newly sampled potential function. 
Many of the small field models from Measure B that produce enough inflation will be initialized very near the maximum, where fluctuations send Hubble volumes into both conjoining half-basins.
Nonetheless, we consider draws of the potential and initial conditions as belonging to one of two classes for the purpose of the analysis below: those initialized in either the true- (Minkowski) or false-vacuum (de Sitter) basins of attraction.\footnote{Recall that the true vacuum basin is the one in which we have artificially shifted an already-low vacuum energy to precisely $\rho_\Lambda = 0$; a false-vacuum basin is one of the two adjacent to the true-vacuum basin, in which the vacuum energy is positive.
We only consider one transition event.}

For Measure B we do not account for inflation in a slow roll interval that is not contiguous with the initial field value.
This is because those inflating intervals are already included in the domain of Measure B, and a history in which slow roll starts higher on the potential and begins to inflate at the top of that region does not differ observationally from a history in which the field is initialized in that interval.
Furthermore, it is less clear cut than in Measure A how we should sample the initial field velocity to make the determination of whether the dynamics quickly reduce to slow roll at a lower site on the potential.

\begin{figure}[t]
	\centering
	\includegraphics[height=5.5cm]{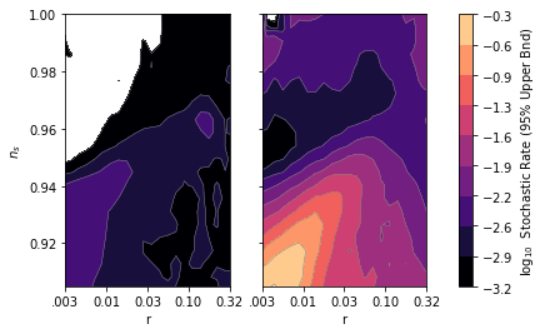} %
	\caption[Stochastic inflation in Measure B, all scales of scalar curvature perturbation amplitude]{95\% confidence upper bound on the rate of incidence of stochastic inflation in Measure B, binned with respect to scalar tilt and tensor-to-scalar ratio, with at least $70$ $e$-folds (left) and 200 $e$-folds (right), subject to epektacratic field scale weighting.}
	\label{fig:276}
\end{figure}

\subsubsection{Initialized in a True-Vacuum Basin}

\paragraph{Stochastic Eternality}
%
In the lefthand plot in Figure~\ref{fig:276}, we depict rates of incidence of stochastic eternal inflation among models with greater than 70 $e$-folds, using epektacratic field scale wighting and binning with respect to the scalar spectral index and tensor-to-scalar ratio.  
Stochastic inflation is most prevalent in the vicinity of $n_s \approx 0.963$ and $r \approx 0.15$;
this corresponds to the quadratic limiting behavior in large field models, where the field excursion during a Hubble time goes like $m_{\text{h}}^{-1}$ making stochastic inflation more likely.
Even then, fewer than 1\% of models from all mass scales in the population are stochastically eternal. 

Rates of stochastic eternal inflation presented here may be suppressed because we require merely that $\epsilon_V,\abs{\eta_V} < 1$ to assume slow roll takes place in the full dynamical evolution, rather than the strong versions of those inequalities.
Although the attractor behavior of the full equations of motion leads to authentic slow roll, this choice could be letting through models that accrue $e$-folds in intervals with larger $V'(\varphi)$ or $V''(\varphi)$ than would be admissible if the strong inequalities were enforced at the initial field value.
If we turn our attention to regions of parameters space in which the scalar index $n_s \approx 1 - 6\epsilon_V + 2\eta_V$ is close to 1, it suggests that the potential slow roll parameters are small enough to satisfy the strong inequality at least near the horizon exit scale.
To compensate for this possible bias and to showcase dependence of the results on $e$-fold count, we also show rates conditioned on at least 200 $e$-folds in the plot on the right of Figure~\ref{fig:276}.

In our region of parameter space around $n_s = 0.96$ and $r \lesssim 0.7$, the effect of requiring more $e$-folds is to only slightly increase the incidence rate of stochastic eternal inflation.
All of the dramatic effects of conditioning on more inflation occur for redder scalar spectra than are viable based on {\it Planck} data.
Observables $n_s$ and $r$ are highly correlated along contours of equal probability of eternal inflation in the lower left region of both plots, corresponding to red spectra and small tensor perturbations.
This can be understood in terms of both quantities' dependence on $\epsilon_V$, which can be made to appear in the stochastic inflation criterion \eqref{seic}.
It seems that before conditioning on the amplitude of the scalar spectrum, our region of this parameter space represents a local {\it minimum} for probability of eternal inflation in this range.

\begin{figure}[t]
	\centering
	\includegraphics[height=5.5cm]{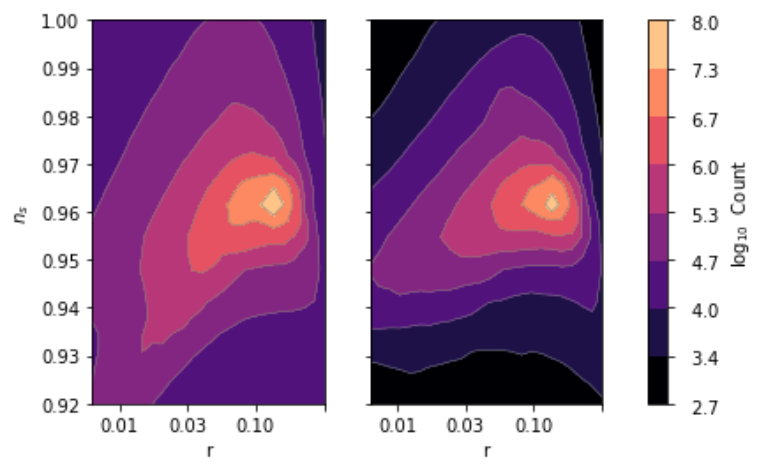}\hspace{.09cm}
	\includegraphics[height=5.5cm]{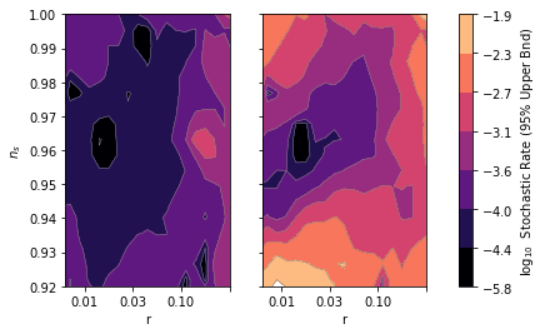}
	\caption[Stochastic inflation in Measure B, small scalar curvature perturbation amplitude]{(Top) Total number of successful models with $Q_s < 10^{-3}$ in each bin.  (Bottom) 95\% confidence upper bound on the rate of incidence of stochastic inflation in Measure B, binned with respect to scalar tilt and tensor-to-scalar ratio, with at least $55$ $e$-folds (left) and 200 $e$-folds (right), for $Q_s < 10^{-3}$ and subject to epektacratic field scale weighting.}
	\label{fig:278}
\end{figure}

Figure~\ref{fig:278} depicts the rates of incidence of stochastic eternal inflation among successful models with small scalar amplitude $Q_s \leq 10^{-3}$ (our maximum likelihood value is close to $10^{-4.3}$), as well as the number of models aggregated in each bin---to inform where sample size is determining the estimate.
Eternal inflation is suppressed considerably for scalar-dominated spectra after conditioning on small scalar perturbations, with rates smaller by up to 3 orders of magnitude compared to the sample of models that merely produce enough inflationary $e$-folds.
Our region of parameter space continues to appear as near a local minimum for the probability of stochastic eternal inflation.

\paragraph{Topological Eternality}


\begin{figure}[t]
	\centering
	\includegraphics[width=12.5cm]{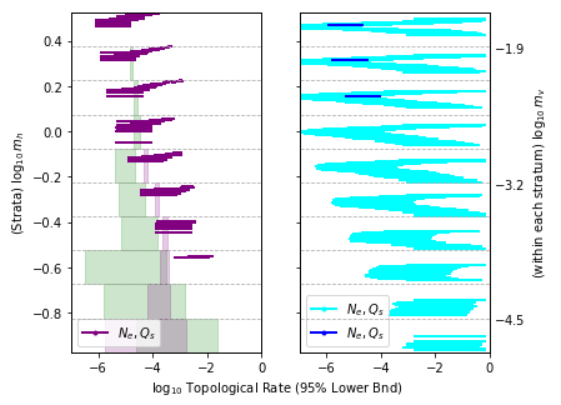}
	\caption[Topological inflation in Measure B]{
		95\% confidence lower bound on incidence rate of models with high probability of topological inflation, among models in Measure A with 70+ $e$-folds.
		In these models, quantum fluctuations are comparable in size to $\abs{\varphi_0 - \varphi_{\text{max}}}$, allowing $\gtrsim 1$ Hubble volume to descend toward the opposite local minimum after a Hubble time with high probability, and produce a persisting topological defect.
		Each line represents one batch of simulations with particular $m_v,m_{\text{h}}$ (only showing batches with at least one positive event per sample or a sample size of 100).  All models within a vertical stratum have the same value of $m_h$ (the center of the stratum on the left axis); vertical position within the stratum reflects $\log_{10} m_v$ (range shown on the right axis).
		In the left plot, green data points are derived from samples conditioned on $n_s$ and $\alpha$; the purple 90\% confidence intervals are conditioned only on minimal $e$-folds; and the shaded bars indicate 90\% confidence intervals taking models from all values of $m_v$ as belonging to one sample. 
		In the right plot, the blue data points are derived from samples of successful models conditioned further on $Q_s$. Darker points have at least one non-stochastic model in the sample, whereas lighter points are determined only by sample size. }
	\label{fig:scatter_207}
\end{figure}


Topological eternal inflation can come about in Measure B, if the field value is initialized close enough to the maximum that fluctuations are likely to result in at least one Hubble volume with a field value on the other side of the barrier.
In Figure \ref{fig:scatter_207}, we depict bounds on the incidence rate---among successful models and those conditioned further on spectral features---of those in which the uniformly sampled initial field value lands close enough to the maximum that fluctuations are likely to result in at least one Hubble volume on the other side of the barrier after a Hubble time.
%
In fewer than 1 in $10^4$ models with a small scalar amplitude do we find this condition to hold.
The only sample in which we get positive events is that of all models with $\mathcal N_e > 70$; in that sample, the incidence of topological eternal inflation does exhibit a dependence on $m_v$, with higher potential scales yielding higher rates of topological inflation.
This follows from the fact the the size of fluctuations goes like $m_v^2$, making it easier to attain large fluctuations that carry the field over the peak.
No events were observed in the population further conditioned on a viable scalar spectral index; the bounds in that sample are determined entirely by sample size.

Figure \ref{fig:scatter_207} does not account for models that are initialized in a stochastic inflation interval contiguous with the maximum, but in which fluctuations from the initial field value are not likely to reach all the way to the maximum after one elapsed Hubble time.  
In such models, the field would gradually climb the potential as it undergoes stochastic eternal inflation, eventually to reach the peak and descend down the other side---becoming also topologically eternal.

\subsubsection{Initialized in a False-Vacuum Basin}

When $\varphi$ lands in a basin with a positive vacuum energy adjacent to the Minkowski basin, we assume that large regions come to occupy the false vacuum.  
We compute the Coleman-de Luccia instanton profile interpolating between the starting basin and ``true'' vacuum\footnote{Since the domain of the Gaussian random field is infinite, one can always find a lower energy vacuum.  We limit consideration to the vacua to either side of the Minkowski basin.}, or determine that a CDL solution does not exist.  
When accounting for gravity, the CDL instanton terminates on the slopes of the barrier rather than precisely at the local minima; so if a solution exists we then initialize $\varphi$ with a new starting position at its terminus on the slope on the true-vacuum side of the barrier, and tally $e$-folds below that point.
If a CDL instanton does not exist, then the Hawking-Moss instanton gives the largest contribution to the transition amplitude between de Sitter and Minkowski basins.
This happens when the top of the barrier is sufficiently flat \cite{A07transitionsbetween}
\begin{equation} V(\varphi_{\text{top}}) \gtrsim \subtext m P^2 \, V''(\varphi_{\text{top}}) \label{eq:CDL_cutoff} \end{equation}
in which case 
 $\varphi$ following a Euclidean classical trajectory either cannot build up enough kinetic energy to close the bubble on the false-vacuum side or loses it to friction.
Well above the Planck scale $m_{\text{h}} = 1$, it is therefore safe to assume that the transition is Hawking-Moss.  In that case, $\varphi$ is re-initialized at the top of the barrier, as in Measure A.

Having landed in the false vacuum basin, {\it generically eternal} would mean that among models that produce observables consistent with {\it Planck} spectral fit and $\rho_\Lambda = 0$ after tunneling, the transition rate is generically below the threshold given in \eqref{fv_rate} or generically Hawking-Moss (leading to stochastic and topological inflation irrespective of the tunneling rate; there is always inflation at the peak when HM dominates).

\begin{figure}[t]
	\centering
	\includegraphics[width=9.6cm]{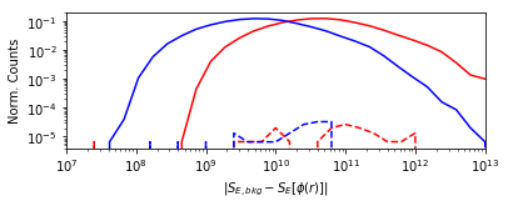} 
	\caption[Distributions of tunneling rate suppression for CDL transitions in Measure B.]{The bimodal distribution of $S_{\text{E,bkg}}-S_{\text{E}}[\phi(r)]$ for Coleman-de Luccia transitions in models initialized in the false vacuum, for $m_v = 0.0025$ (red) and $0.0042$ (blue). The normalized counts for values of the tunneling suppression are plotted with respect to its absolute value;  the solid lines correspond to slow tunneling, for which inflation is eternal, while the dashed lines are fast tunneling.}
	\label{cdl_rates}
\end{figure}

\paragraph{Coleman-de Luccia}

When CDL instantons exist, 
transition rates generically fall below the eternal inflation upper bound, as shown in Figure~\ref{cdl_rates}.
The distributions for different energy scales of inflation differ by a translation in the log domain, scaling with $m_v^4$. Vertical mass scales on the order $m_v^4 = \mathcal O(10^{-5})$ would correspond to a point where the distribution has support in the vicinity of $9/4\pi$ and the determination of genericity becomes more nuanced, but that is far above the range where small scalar and much smaller tensor curvature perturbations are likely to be found at horizon exit. 
So models with a CDL transition are generically eternal within the scope of this analysis.
The challenge is to get enough $e$-folds on the other side of the barrier for a CDL transition to precede inflation, and to characterize statistical behavior of the tunneling rate among those very rare events.

\begin{figure}[t]
\centering
\includegraphics[width=8cm]{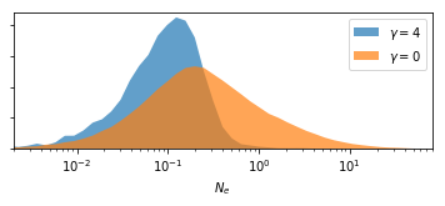} 
\includegraphics[width=9.6cm]{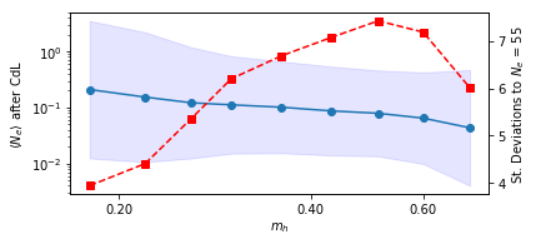} \hspace{.3cm} 
\caption[Distribution of number of inflationary $e$-folds after CDL transition]{(Top) Distributions of the number of slow roll $e$-folds in the true vacuum basin after CDL tunneling, for two values of the shape parameter $\gamma$ characterizing the potential in \eqref{grf_cov_var}.  (Bottom) Moments of the distribution of number of $e$-folds after CDL tunneling when inflation ends in the tunneled-to basin, as a function of $m_{\text{h}}$.  The blue plot (solid line with circular markers, left axis) shows the mean of $\log_{10} N_e$ and the shaded 2-$\sigma$ confidence interval.  The red (dashed line with square markers, right axis) is the number of standard deviations between the mean and $\log_{10} 55$.}
\label{fig:efoldsaftercdl}
\end{figure}

The distributions of the number of $e$-folds after Coleman-de Luccia tunneling for $\gamma = 0$ and $\gamma = 4$ (referring to the shape parameter in \eqref{grf_cov_var}), accounting for all field scales with epektacratic weighting, are shown in Figure~\ref{fig:efoldsaftercdl}.
Among models aggregated from all field scales $m_{\text{h}}$ and for which the field value after tunneling satisfies slow roll, the distribution resembles log-normal for $\gamma=0$ in \eqref{grf_cov_var}, with an expectation value of less than one $e$-fold (in effect, no inflation).
Also shown are the expectation values and 2-$\sigma$ ranges (assuming log-normal) for populations sampled with a single field scale and $\gamma=0$, along with the number of standard deviations between the mean and 55 $e$-folds where horizon exit of CMB modes could occur.
For $\gamma = 4$, practically the entire distribution is localized below 1 $e$-fold---yielding no inflation post-transition.

Determining the rate of non-eternal inflation comes down to the distribution of the tunneling rate among models with CDL instanton solutions that are just on the threshold of not existing---with the field landing very close to the maximum, but outside the stochastic inflation regime around the peak.
For this reason, we consider models in which the CDL solutions dominate as effectively not contributing to the population of observationally viable models, for the purpose of determining whether eternality is generic under Measure B.
For a similarly defined measure in which initial field values in the true vacuum basin are excluded, one would need a way of sampling such very rare potential shapes that give sufficient inflation after a CDL tunneling event, in order to characterize the prevalence of eternal inflation among small field models.


 \paragraph{Hawking-Moss}
%
We take the Hawking-Moss calculation to give the rate at which Hubble volumes occupying the false vacuum basin {thermally} fluctuate into the true-vacuum basin, with energy comparable to the height of the barrier \cite{A07transitionsbetween}.  
Since the potential is only sampled at the false vacuum and at the top of the barrier, the Hawking-Moss transition rate is independent of $m_{\text{h}}$; so we can expect this same distribution at higher field scales as well, in regions where $Q_s$ is likely to match observation.
Tunneling rates only begin to approach the fast-tunneling regime when the potential approaches the Planck scale.
This corresponds to an enormous scale for the inflaton $m_{\text{h}} \sim 10^6$ in order to get a small enough scalar amplitude.

The question of whether eternal inflation can be avoided then comes down to how we interpret the fast-tunneling Hawking-Moss instanton.  It is not eternal on the usual false-vacuum grounds, but supposedly ends with the field everywhere in a Hubble sized region sitting atop the maximum, where we would expect that it would fluctuate away from the peak into the true vacuum basin and give topological inflation.


%

\subsection{Measure C: Hilltops}

Suppose the inflaton field value in an initial Hubble volume is drawn at some characteristic distance in field space---say $H/2\pi$, or one standard deviation for Hubble-scale fluctuations---away from the local maxima of potential barriers randomly sampled by the procedure defining Measure A.  
If the field space interval around the maximum in which fluctuations dominate is narrower than this gap, then the model has a chance to avoid stochastic and topological inflation.
%
%
How often is stochastic eternal inflation localized entirely within that neighborhood of the peak, with enough inflation lower on the potential to solve the horizon problem? 
In principle, some Measure A models with inflation at the peak can be excluded from Measure C, if fluctuations are larger than the inflating interval around the maximum that includes the would-be horizon exit scale.

%
%
%

\begin{figure}[t]
\centering
\parbox{\textwidth}{
\includegraphics[height=4.5cm]{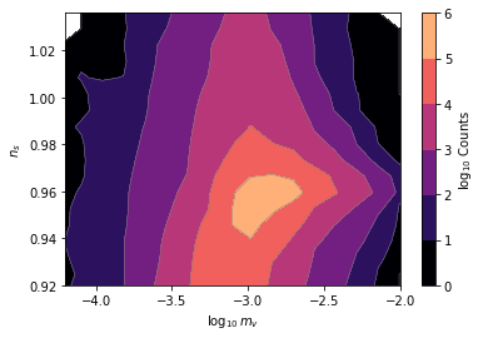}
\includegraphics[height=5.5cm]{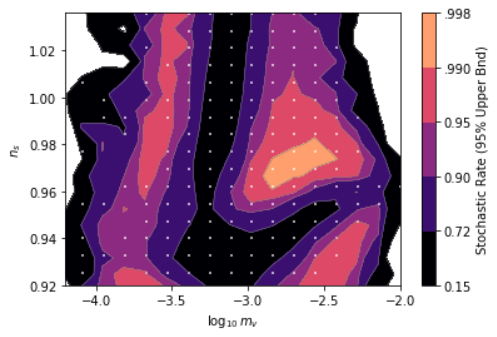} } 
\parbox{\textwidth}{
\includegraphics[height=4.5cm]{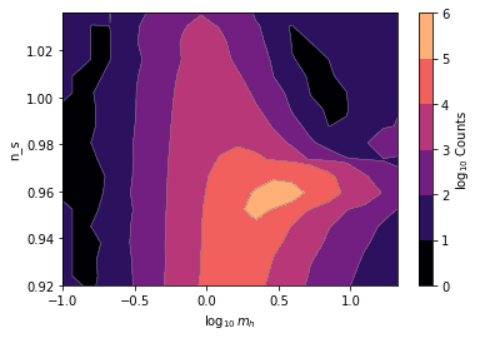}
\includegraphics[height=5.5cm]{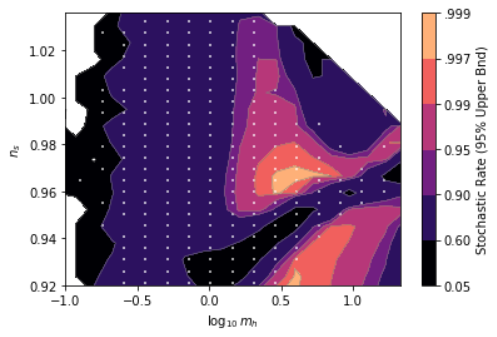} 
}
\caption[Stochastic inflation in Measure C, binned with respect to $n_s$ and either $m_v$ or $m_{\text{h}}$.]{
(Left) Bin counts. (Right)
95\% confidence lower bound
on the rate of incidence of models with fluctuations smaller than the width of the stochastic inflation interval around the maximum, under epektacratic weighting in Measure C, binned with respect to scalar tilt and (top) vertical or (bottom) horizontal mass scale, and conditioned on $r < 0.064$, and $-4.5 < \log_{10} Q < -4.1$.
The grid points indicate the centers of bins with a non-zero number of \emph{non}-eternal models.  Where grid points are absent, the reported bound is determined only by sample size; a small, uninformative lower bound in regions with few samples.}
\label{fig:contour_070}
\end{figure}


A large field widens the fluctuation-dominated interval near the peak, but also correlates with tall peaks in regions of model space in which scalar curvature perturbations reflected in the CMB are the right size; the latter leads to larger fluctuations around the peak.
Approximating as quadratic the neighborhood of the potential around the maximum in which fluctuations dominate, we can get a sense for when quantum fluctuations are larger than the width of that stochastically inflating interval.
Evaluating Eq.~\eqref{seic} at a field value separated from the maximum by the width of Gaussian fluctuations $\delta\varphi$, we have
\[ (V(\varphi_{\text{peak}}) + \tfrac1{2} V''(\varphi_{\text{peak}}) \, \delta\varphi^2)^{3/2} > 6.6 \, \abs{V''(\varphi_{\text{peak}}) \, \delta\varphi} M_{\text{P}}^3 \]
Taking quantum fluctuations of size $\delta\varphi^2 = H^2/4\pi^2 = \tfrac2{3\pi} m_v^4 \subtext m P^2 f(x)$, where $x$ is the dimensionless field value at the peak,
\[ m_v^6 \left( f(x) - \frac1{3 \pi} \frac{m_v^4}{m_{\text{h}}^2} f(x) f''(x) \right)^{3/2} \gtrsim \frac{6.6}{16\pi^2 \sqrt{3}} \frac{m_v^6}{m_{\text{h}}^2} f''(x) \sqrt{f(x)} \]
 and assuming typical values for the shape function of order unity, $f(x) = f''(x) = 1$, we obtain in terms of the mass scales characterizing the potential:
\[ \left( 1 - \frac1{3 \pi} \frac{m_v^4}{m_{\text{h}}^2} \right)^{3/2} \gtrsim \frac{6.6 \, m_{\text{h}}^{-2}}{16\pi^2 \sqrt{3}} \]
Considering the large field regime, we set $m_{\text{h}} \sim 10^6\, m_v^2$ to delineate the region in Figure \ref{fig:142_Q} where the scalar amplitude $Q_s$ most often takes its value modeled from measured data.
We find that 1-$\sigma$ fluctuations deposit the field beyond the breakdown of the stochastic inflation criterion when $m_v \lesssim 10^{-3.4}$ or $m_h \lesssim 1$---just about where the highly localized linear trend begins in Figure \ref{fig:142_Q}.

So granted the above assumptions we should expect a comfortably wide site of stochastic inflation at the peak, among large field models that are most likely to give the observed amplitude of scalar curvature perturbations.
Then, what explains the hook-shaped incursion of the contours of relatively low incidence of peaky stochastic inflation in Figure~\ref{fig:contour_070}, approaching $n_s \sim 0.96$ from below in the large-$m_v$ and large-$m_h$ regions?
These plots show the rate of incidence of stochastic inflation among Measure C models respecting {\it Planck 2018} constraints on the scalar and tensor amplitudes, binned with respect to the scalar index and the vertical or horizontal mass scale.
%

Large-field models producing a tensor-to-scalar ratio less than 0.07 are those in which horizon exit occurs far from the quadratic minimum where $\epsilon_V$ is small; for the scalar spectral index to also be more red than the quadratic limit of 0.96, the second potential slow roll parameter $\eta_V$ must be larger at horizon exit, correlating with a narrow stochastic inflating interval around the peak. 
Inflation yielding a redder spectrum with small scalar and tensor perturbations thus typically breaks the approximations used in the above calculation, with $f''(x)$ atypically large.
Entry of higher order terms that would break the quadratic approximation coincides with breakdown of \eqref{seic}.
(Taking into account that the field excursion is in fact greater than \eqref{seic} assumes when $V''(\varphi)$ is significant, the stochastic eternal interval is actually smaller; so the above estimate of the lower bound on $m_v$ is conservative---more cheritable to stochastic inflation.)
In the intermediate zone $-3.5 < \log_{10} m_v < -3$, inflation no longer necessarily ends in the quadratic regime, but peaks that have atypically high energy can, and they have a better chance of scoring a selection boost.
As the energy scale shrinks further, without much new selection pressure on $m_{\text{h}}$ coming from the $Q_s$ constraint, fluctuations once again become smaller than the eternal inflation interval, and stochastic inflation becomes more prevalent below $\log_{10} m_v \sim -3.5$.
Below $m_v \sim 10^{-4}$ or $m_{\text{h}} \sim 0.2$, most models with small $Q_s$ are far too red, 
and we lack sufficient simulated data in the range of Figure \ref{fig:contour_070} to place bounds.
When binning with respect to field scale, incidence rates for stochastic inflation are consistently low below $m_h \sim 1$---in the realm of \emph{common} but not \emph{generic}.




\begin{figure}[t]
\centering
\includegraphics[height=5.5cm]{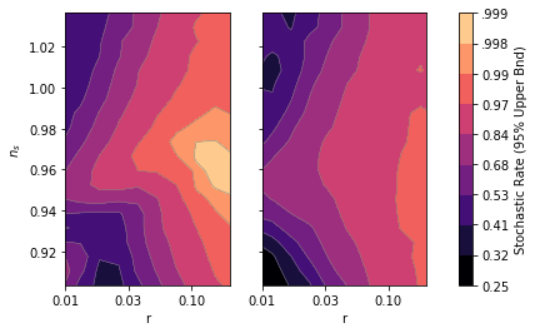} 
\caption[Stochastic inflation in Measure C, binned with respect to $n_s$ and $r$.]{95\% confidence lower bound on the rate of incidence of stochastic inflation in Measure A, binned with respect to scalar tilt and tensor-to-scalar ratio, and conditioned on $-4.5 < \log_{10} Q_s < -4.1$.  The left plot includes all mass scales with epektacratic weighting.  The right includes only $m_{\text{h}} \leq 1$.}
\label{fig:contour_138}
\end{figure}


So for the range $0.955 < n_s < 0.975$, eternal inflation peeks into the realm of genericity (at $ > 95\%$ incidence) by the stochastic mode for vertical mass scales in the middle of our range. 
Figure \ref{fig:contour_138} shows rates of stochastic inflation conditioned on $-4.5 < \log_{10} Q < -4.1$, and binned with respect to scalar spectral index and tensor-to-scalar ratio, including models from all mass ranges represented in Figure \ref{fig:142_Q} with epektacratic weighting.
Stochastic inflation is generic for all spectral index values when $r > 0.1$, including all mass scales or just those less than $m_{\text{P}}$.
For small tensor-to-scalar ratio, eternal inflation is suppressed.


\section{Concluding Remarks}
\label{sec:discussion}

Clearly the question ``Is inflation generically eternal?'' cannot at present have a single definitive answer, due to the ambiguity in the choice of a measure over viable inflationary cosmologies that may or may not be eternal.
Furthermore, it is unclear whether it is even sensible to seek a single verdict for any particular measure, akin to the {\it top-down} probabilities discussed in \cite{PhysRevD.28.2960}.
Our framing offers a bit more latitude for exposition: ``By what modes and to what extent is inflation generically eternal with respect to some simple measures?''  

To address these questions, we implemented measures in which the potential is a Gaussian random field and the inflaton is initialized according to either a uniform distribution (B), a weighted sampling of local maxima (A), or a weighted sampling of field values displaced $H_{\text{max}}/2\pi$ from local maxima (C).
We simulated inflationary histories with potentials and initial conditions drawn from these measures, and analyzed rates of occurrence of the three modes of eternal inflation among subpopulations of models conditioned on matching various observables from the \textit{Planck 2018} survey of the Cosmic Microwave Background.

We found that topological eternal inflation is only clearly generic when initializing at the top of the barrier, an initial state from which it is nearly inevitable that an inhomogeneous configuration interpolating across the barrier will come about.
Stochastic inflation is generic at large field scales in Measures A, and in pockets of the space of mass scale parameters in Measure C; it is generically absent in all regions of the space of CMB spectral parameters examined among models conditioned on minimal inflation.  The simulated data suggest that transitions between de Sitter minima in viable cosmological histories---whether Coleman-de Luccia or Hawking-Moss---generically result in eternal inflation.



\paragraph{Further Research}

We were unable to address the typicality of eternal inflation at small field scales---less than an order of magnitude below the Planck mass.
One could take an importance sampling approach to probe the measure over viable small field models and very rare false vacuum transitions, rather than sampling the measure's native distribution and waiting low-probability samples to appear by chance.


In higher dimensional potentials spanned by multiple scalar fields, there can be a reduction in the effective field space dimension experienced by a  trajectory as it approaches a minimum.  
For particular multfield models with a lot of extra symmetry, multifield dynamics reduce to effectively single-field well before the era of inflation that influences what we observe.
Does a measure on initial conditions naturally emerge given statistical characteristics of the landscape, identifying where on the slope the potential becomes effectively one-dimensional?
The CosmoTransitions package includes code for computing Euclidean action-minimizing trajectories in multifield potentials, and could be used to investigate false vacuum inflation in $N$-D Gaussian random fields.  

\bibliography{Reference/main}{}
\bibliographystyle{unsrt}

\appendix
\section{Monte Carlo Methods Continued}
\label{app:mc_methods2}

\subsection{Simulation Design}
\label{app:sim_design}
For each instance toward building up the distribution, the steps are as follows:


\begin{enumerate}

\item \textbf{Initialization}
\begin{enumerate}
    \item Sample a 1-D Gaussian random field $f(x)$ according to \eqref{grf}. From this and constants $m_v,\,m_h$ construct the potential $V(\varphi)$ according to \eqref{eq:grfV}.
    \item Initialize the inflaton at $\varphi_{\text{start}}$ according to one of the Measures A, B, or C outlined above.  
    \item Determine the potential energy $\rho_\Lambda$ of the minimum of the starting basin and in one neighboring basin in both directions.  (For Measure A, the starting basin adjacent to the initial peak is chosen randomly weighted by width.)  If $\abs{\rho_\Lambda}$ in any basin in this search space is below a threshold, shift the potential so that $\rho_\Lambda = 0$ in that basin. If $\rho_\Lambda$ is negative and less than this threshold in the starting basin, abort.
\end{enumerate}

\item \textbf{Instanton Pre-selection}

Computing instanton profiles is time consuming, so we take the following steps to determine if a tunneling event is likely to be followed by sufficient inflation to produce a possibly observable universe in another basin of the potential.  
\begin{enumerate}
\item If initialized in the true vacuum with $\rho_\Lambda=0$, continue to (\ref{charrho}).
\item If the thin-wall or Hawking-Moss approximations hold, continue to (\ref{tunnel_check}).
\item Taking the cutoff CDL instanton terminus on the true-vacuum side $\varphi_{\text{edge}}$ to coincide with $ V(\varphi_{\text{edge}}) = 0.05\, V_{T} + 0.95 \, V_{\text{bar}}$, compute the maximum number of $e$-folds of inflation accrued over any field space interval in which the potential slow roll conditions are met between $\varphi_{\text{edge}}$ and the true minimum.
If the maximum $e$-fold count is less than 70, abort.
\end{enumerate}

\item \textbf{Check for Quantum Tunneling} 
\label{tunnel_check}
\begin{enumerate}
\item If the thin-wall approximation is strongly valid or $m_h \gg m_{\text P}$ (Hawking-Moss eminent), compute the transition rate, otherwise
\item Compute the Coleman-de Luccia tunneling profile; determine the instanton terminus on the true-vacuum side; compute the number of $e$-folds of inflation, assuming inflation takes place anywhere below the terminus where the potential slow roll conditions are weakly met ($\epsilon_V,\eta_V < 1)$.
\end{enumerate}

\item \textbf{Characterize Slow Roll}
\label{charrho}
\begin{enumerate}
    
    
    \item Look downhill from $\varphi_{\text{start}}$ for breakdown of the slow roll approximation, $\varphi_{\text{end}}$.  
    
    \item Compute the number of $e$-folds $\mathcal N_e$ in the current basin.  If $\mathcal N_e < 70$, skip to (\ref{datacol}).
    
    \item Find $\varphi_{\text{exit}}$, the field value at the horizon exit scale for CMB fluctuations, taken to be 55 $e$-folds before the end of inflation.\footnote{Our $e$-fold cutoffs (70 for successful inflation, 55 for imprinting of CMB fluctuations) of course depend on the fiducial reheating model.  One could include those models in the input space, but we opt not to include that freedom in this analysis as doing so would likely obscure our conclusions.}
    
\end{enumerate}

\item \textbf{Check for Eternal Inflation}
\label{eicheck}
\begin{enumerate}

    \item
    Evaluate the stochastic inflation criterion \eqref{seic} between $\varphi_{\text{start}}$ and $\varphi_{\text{end}}$ in each basin.  
    

    \item 
    
    Check the second potential slow roll condition at all local maxima along the trajectory; compare to the upper bound for topological inflation (see \ref{app:topol}).
    
    \item If a transition into the basin with $\rho_\Lambda=0$ is followed by enough $e$-folds,  compute the transition rate $\lambda$ and compare to the upper bound in Eq.~\eqref{fv_rate}.
    
\end{enumerate}

\item \textbf{Data Collection}
\label{datacol}

Record observables if inflation ends with $\mathcal N_e > 70$ in a vacuum with $\rho_\Lambda \ll 1$, along with indicators for eternal inflation:
\begin{align} 
\mathbf p_{\text{obs}} &= (Q_s,r,n_s,\alpha,n_t,\delta\rho/\rho,\log \abs{\Omega-1}) \label{eq:paramvecapp} \\
\mathbf p_{\text{eternal}} &= (N_s,\langle\mathcal  N_{e,\text{stoch}}\rangle,b_t,\lambda_{\text{fv}},H_{\text{F}},b_{\text{HM}}) \label{eq:paramvec2app}
 \end{align}

\end{enumerate}


The parameters in Eq.~\eqref{eq:paramvecapp} are the same as those defined in \S~\ref{sec:simflow}.

\paragraph{Criterion for Inflation Discontiguous with an Initial Peak}

To determine if $\dot\varphi$ comes in below the slow roll attractor bound \eqref{slowrollattractor} at the high potential energy end of a field space interval in which $\epsilon_V,\eta_V$ fall below 1, we integrate the coupled equations of motion for the homogeneous scalar field and the metric (only the scale factor $a(t)$)
\begin{align} \ddot\varphi &= -3 H(a,\dot a) \dot\varphi - V'(\varphi) \\
\ddot a &= 8\pi G a \left( \dot\varphi^2 - V(\varphi) \right)
\end{align}
backward in time, with final conditions $\varphi = \varphi_{\text{sr}}$, $\dot\varphi = \mathrm{sgn}(\varphi_{\text{sr}}-\varphi_0)\times\sqrt{V(\varphi_{\text{sr}})}$.

If the solution overshoots the peak in the past, it would mean a velocity in the direction of $\varphi_{\text{sr}}$ at some finite initial time.
If the system were conservative, we could safely assume that as we dial that initial velocity at the peak down to zero, $\dot\varphi$ approaches a value less than $\sqrt{V(\varphi)}$ at $\varphi_{\text{sr}}$.  The presence of Hubble friction complicates things somewhat, as it becomes possible that reducing this initial field velocity also reduces friction to the point that one gets a \emph{greater} field velocity where slow roll begins.  
This can be tackled iteratively by a method of overshoot-undershoot, integrating backward in time trying to land with $\varphi$ atop the peak at $t \to -\infty$.
More expeditiously, in the case of an overshoot we can then initialize with a small field velocity at the peak in the direction of the true-vacuum basin, and assume nothing changes as that small velocity vanishes.

\begin{figure}[t]
	\centering
	\includegraphics[width=7cm]{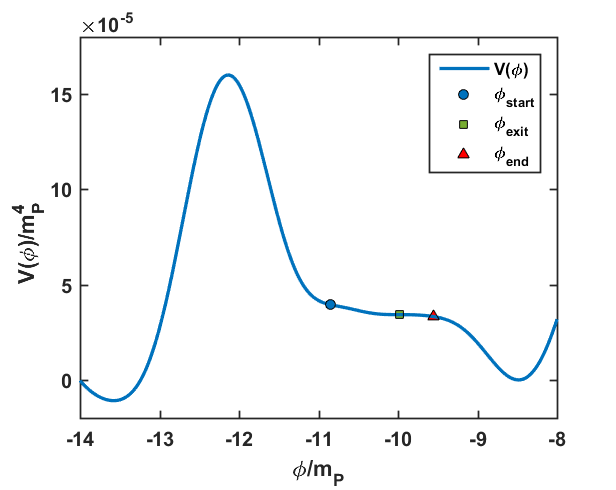}
	\includegraphics[width=7cm]{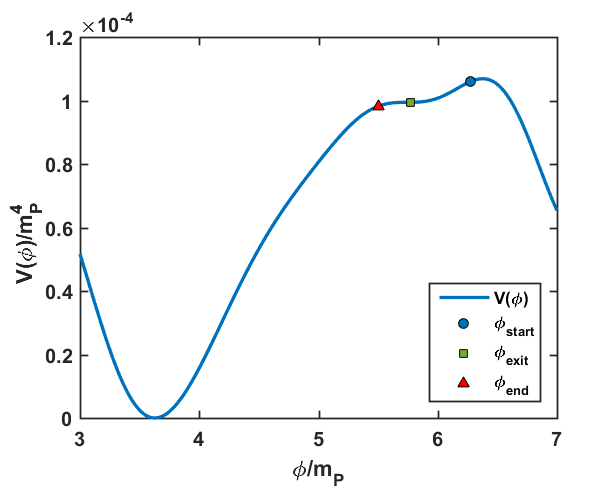}
	\caption[Examples of delayed inflation in Measure A]{Examples of models drawn from Measure A that fall into subset $D$ (as defined in \S~\ref{sec:MeasureA}).  The potential on the left represents the spirit of successful inflation taking place in field space interval discontiguous with an initial non-inflating peak.  The potential on the right has $\eta_V < -4/3$ at the peak, but the curvature quickly shrinks to within the slow roll attractor; initialized with a small field velocity, inflation really continues uninterrupted between the peak and the interval in which the potential slow roll conditions are met, with the kinetic energy never rivaling $V(\phi)$. Both models are treated the same in our simulations, raising the question of how many such models are actually free of stochastic eternal inflation.}
	\label{delayed_inflation_examples}
\end{figure}


\subsection{Instanton Computation}

Treating transitions between de Sitter minima, we require the instanton profile $(\varphi(\xi),\rho(\xi))$ to determine where the field is to be initialized on the true-vacuum side of the barrier after the transition.
We also need the transition rate to compare the rates of bubble nucleation or stochastic ascent of the peak against the expansion rate.

\begin{figure}[t]
\centering
\includegraphics[width=8.5cm]{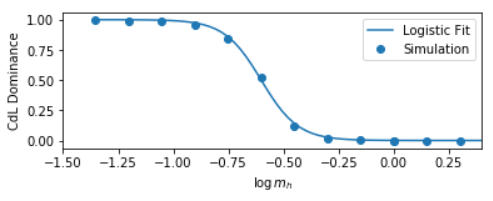} 
\caption[Coleman-de Luccia or Hawking-Moss dominance vs. field scale]{Fraction of simulated models in which a Coleman-de Luccia instanton solution exists. At larger horizontal mass scales (flatter potential peaks), the Hawking-Moss instanton is dominant.}
\label{fig:cdl_dominance}
\end{figure}

\paragraph{Instanton Pre-selection}
As discussed in \S~\ref{fvei} and illustrated in Figure~\ref{fig:cdl_dominance} from simulated data,
the Hawking-Moss instanton dominates the transition between de Sitter minima if the top of the barrier is sufficiently flat \cite{A07transitionsbetween}:
\begin{equation} V(\varphi) / m_{\text{Pl}}^2 \gg V''(\varphi) \implies m_h \gg (8\pi)^{-1/2} \label{eq:cdl_cutoff} \end{equation}
%
Well above the scale $m_h = (8\pi)^{-1/2}$, it is safe to assume that the transition is Hawking-Moss; so the transition rate is closed-form and easy to compute.
At field scales much smaller than $M_P$, we enter the thin-wall regime in which the CDL instanton dominates. 
The thin-wall instanton tends to terminate very near the true vacuum, traversing the barrier over a small interval in the Euclidean radial coordinate relative to $H_{\text{F}}^{-1}$.

In the intermediate regime around $m_h = (8\pi)^{-1/2}$, it may not be an easy determination which instanton contributes most to the transition; we must compute the profile.
In order to reduce program time allocated to computing transition rates in this regime, we first perform a check that inflation can start and end in the adjacent basin and that the maximum amount of inflation likely to occur is sufficient to produce $\mathcal N_e > 70$.  
%
If the effective mass is such that $0.1 < (m_{h,\text{eff}}^2 \equiv V/V_{,\varphi}) < 1$, we determine $\varphi_{\text{edge}}$ such that \[V(\varphi_{\text{edge}}) = 0.05\, V_{T} + 0.95\, V_{\text{bar}} \] 
and take it as our trial starting point in the new basin. 
We search down the slope for the start (if the potential slow roll conditions are not already met at $\varphi_{\text{edge}}$) and end of slow roll ($\epsilon_V, \abs{\eta_V} < 1$), and
compute the number of $e$-folds that elapse in that interval assuming slow roll.
If it is greater than the 70 $e$-folds needed to obscure any potentially observable relics of the transition, then we proceed with the full instanton calculation to determine where precisely the field is deposited.  
Otherwise, we assume that a transition does not result in an observationally viable universe, and so does not inform $f_m(\mathbf p \mid \mathbf p_{\text{obs}})$; we discard and continue to the next randomly drawn potential function.
%

\paragraph{Obtaining the Profile}

To determine the action-extremizing instanton profile for a given potential $V(\varphi)$, we use the algorithm employed in the CosmoTransitions package published with \cite{Wainwright:2011kj}, modified to accommodate parallel processing in Matlab.
The algorithm assumes initial conditions $\dot\varphi(0) = \rho(0) = 0$ with $\varphi(0)$ on the true-vacuum side, and takes the endpoint of the trajectory to occur at $(\varphi,\dot\varphi) = (\varphi_{\text{F}},0)$ (came to rest at the false vacuum) or $(\rho,\dot\varphi) = (0,0)$ (geometry closed with no discontinuities).
The former stopping criterion is only possible in the absence of gravity, though it can be approached in the thin-wall limit.
If the geometry closes ($\rho(\xi > 0) = 0$) with $\dot\varphi(\xi) \neq 0$, then that solution is singular and not admissible.
The steps of the algorithm are as follows:


{ \def\baselinestretch{1.}
\begin{enumerate}
\item Guess a starting field value on the true-vacuum side of the barrier.
\item Integrate equations of motion for the scalar field $\varphi(\xi)$ and Euclidean radius $\rho(\xi)$ of the bubble as a function of the radial coordinate $\xi$. \label{step_integrateprofile}

            
\item Stop integrating when one of the follow events occurs:
    \begin{enumerate}
        \item $\varphi(\xi)$ approaches $\varphi_{\text{F}}$ with $\dot\varphi(\xi) \approx 0$ (Converge)
        \item $\dot\varphi(\xi)$ approaches 0 with $\varphi \approx \varphi_{\text{F}}$ or $\dot\rho(\xi) \approx 0$ (Converge)
        \item $\dot\varphi(\xi)$ changes sign with $\varphi \neq \varphi_{\text{F}}$ (Undershoot)
        \item $\varphi(\xi)$ passes $\varphi_{\text{F}}$ (Overshoot)
        \item $\dot\rho(\xi)$ approaches $-1$ with $\dot\varphi(\xi) \approx 0$ (Converge) or $\dot\varphi(\xi) \not\approx 0$ (Overshoot)
        \item $\rho(\xi)$ changes sign (Converge)
    \end{enumerate}
\item If converged, we're done; return the profile.
\item If within a tolerance value of the top the barrier, report a single data point that fully characterizes the Hawking-Moss profile.
\[ \left\{\varphi,\dot\varphi,\rho,\dot\rho,\ddot\rho \right\} \left( \tfrac\pi2 w_{\text{top}}^{-1} \right) = \left\{\varphi_{\text{top}},\,0,\, w_{\text{top}}^{-1},\,0,\,-w_{\text{top}} \vphantom{\dot\varphi} \right\} \]
\item If the integration overshoots, move the guess closer to the maximum;
if it undershoots, move the guess closer to the true minimum. 
\item Go to Step \ref{step_integrateprofile}.
\end{enumerate}
}

\paragraph{Transition Rates with Gravity}
The tunneling rate in terms of the Euclidean action for the bubble and for the de Sitter background is computed as
\begin{equation} \lambda H_{\text{F}}^4 \approx (\tfrac12 \sigma \bar R)^2 \exp( S_{E,\text{bkg}} - S_{E,\text{bubble}}) \label{tunnelingrate} \end{equation}
where we take $\bar R = (R_0+R_1)/2$ for the purpose of computing the prefactor, $\sigma$ is the bubble tension, and $(\tfrac12 \sigma \bar R)^2$ is the approximate (thin-wall) prefactor.
Beyond the outer radius $R_1$, the bubble and the de Sitter background have the same geometry and field configuration, so those contributions cancel out when computing the transition rate.

For thin-wall bubbles, the initial bubble radius $R_0$ defines the whole geometry; for all bubbles, there is a finite radius inside of which $\dot\varphi = 0$.
To compute the transition rate from the profile, we first compute the curvature of the bubble interior
\[ w_{\text{int}} \equiv \sqrt{\tfrac{\kappa}{3} V(\varphi(R_0))} \]
When $w_{\text{int}} = 0$ the geometry of the interior is Minkowski.
(The instanton profile typically terminates at a value $\rho > 0$, where the field velocity $\dx\varphi/\dx\xi$ effectively vanishes.)
The vacuum in the tunneled-to basin is always Minkowski for the simulation settings chosen for this analysis; however, large-$H$ de Sitter bubbles are also supported in the code, and may result when a sharply peaked barrier is adjacent to a flat interval on the potential in which the potential slow roll conditions are satisfied.
When the vacuum energy in the interior is positive, the radius $\rho(\xi)$ of an anulus on the 4-sphere as a function of the distance from the pole goes like
$ \rho(\xi) = w_{\text{int}}^{-1} \sin(w_{\text{int}} \xi) $.
The term contributing to the Euclidean action from the bubble interior are then
\begin{equation} S_{E,\text{int},\varphi} = \int_0^{R_0} \dx \xi \, 2\pi^2 \rho(\xi)^3 \, V(\varphi(0)) = \frac{ 3\mathcal V_{\text{int}}(R_0) \, w_{\text{int}}^{2}}{\kappa} 
 \end{equation}
for the field, and in the de Sitter case
\begin{align*} S_{E,\text{int},\rho} &= \int_0^{R_0} \, \frac{2\pi^2 \, \dx \xi}{\kappa w_{\text{int}}} \sin(w_{\text{int}} \xi) \left( -\sin(w_{\text{int}} \xi)^2 + \cos(w_{\text{int}}\xi)^2 - 1 \right) = -\frac{ 6\mathcal V_{\text{int}}(R_0) \, w_{\text{int}}^{2}}{\kappa} \end{align*}
for the background geometry.
But this is twice the magnitude and opposite in sign to the field contribution, so within the inner radius of the bubble the contribution is equal to $-S_{E,\text{int},\varphi}$.
For Hawking-Moss instantons, the ``bubble interior'' covers the whole compact space, and we leave out the prefactor in \eqref{tunnelingrate} as there is no analogue to a bubble wall to be perturbed in the standard calculation.  Likewise, the background de Sitter configuration consists entirely of the bubble ``exterior.''


With the interior and exterior covered, we add the contribution from the bubble wall where $V(\varphi)$ is not constant and $\rho(\xi)$ takes a different form.  We integrate the full form of the Euclidean action \eqref{full_euc_action} from the inner radius to the outer radius.

\subsection{Statistical Methods}
\label{app:stats_methods}

\paragraph{Mass Scale Weighting Schemes}
When computing statistics, we adopt one of the following weighting schemes to aggregate simulated models sampled from an array of mass scales $m_v$ and $m_h$:
\begin{itemize}
	\item
	The \emph{epektacratic} weighting scheme (rule by expansion) samples an equal number of potentials for each pairing of mass scales $m_v$ and $m_h$, and lets them succeed or fail at producing sufficient $e$-folds of inflation.  The total population is aggregated from successful inflation models at all mass scales, and that population is used to determine rates.  Naturally this scheme will tend to give more representation to large field models.
	%
	\item
	The \emph{democratic} scheme gives every mass pairing within the specified range equal weight in informing $f_m(\mathbf p_{\text{eternal}} \mid \mathbf p_{\text{obs}})$ in \eqref{likelihood}, regardless of how common or rare it is for models comprising each to produce enough inflation.  From each pairing, we sample as many potentials as it takes to get an equal number of successful models, or we give lower-expansion mass pairings extra weight to the same effect.  
\end{itemize}

\paragraph{Rate Estimation}

We would like to adopt something like a uniform prior on the rate of incidence of eternal inflation (denoted $\lambda$) in each bin.
However, it is not obvious whether we should work in terms of $\lambda$ or $\log \lambda$.  (Are we ambivalent with respect to the rate itself, or with respect to its order of magnitude?)
As a middle road, we adopt the Jeffreys uninformative prior $p_J(\lambda)\;\propto\; \mathrm{Beta}(\lambda; \tfrac12, \tfrac12)$ for the rate parameter of the binomial distribution, which is invariant under reparameterization between the linear and log domains of the rate parameter.  
We would then take the maximum likelihood value (using the uninformative prior) as our estimate of the incidence rate.
However, for bins in which the number of models with or without a given mode of eternal inflation is zero, the maximum likelihood rate is 0 or 1, and does not account for information we have from the sample size of that bin.
For this reason, we often report a 95\% confidence upper and/or lower bound on the incidence rate $\lambda$ in our contour plots, which includes sample size information and makes for smooth contours in regions of parameter space in which positive events may be scarce.  (Naturally, in the case of zero positive events the bound is determined entirely by the sample size.)

\begin{figure}[h]
	\centering
	\includegraphics[width=10cm]{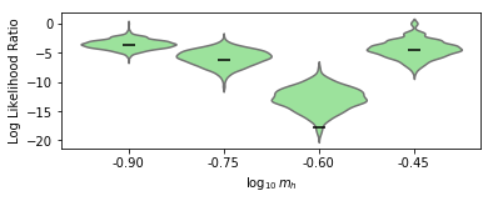}
	\caption[Hypothesis test for independence of non-stochastic delayed inflation of energy scale $m_v$]{Distributions of log likelihood ratios, comparing maximizing over one rate parameter versus independent rate parameters for each $m_v$, from Monte Carlo simulation assuming the former null hypothesis. Ratios for the simulated data are indicated in black.  Only for $m_h = 0.25$, with p-value 0.016, should we consider rejecting the hypothesis of a well defined rate of incidence of delayed non-stochastic inflation independent of $m_v$.}
	\label{violin}
\end{figure}

\paragraph{Testing Scale Invariance of Stochastic Eternal Inflation in Measure A}
It looks as though varying $m_v$ has no effect on the rate of stochastic eternality among Measure A models with successful inflation and those further conditioned on $n_s$ and $\alpha$. 
This would mean that once we have the primary inflation epoch starting on a ``shelf'' below the peak (as depicted in Figure~\ref{delayed_inflation_examples}), it is never borderline between stochastically eternal and non-eternal; if it were, then increasing $m_v$ while holding $m_h$ constant would have the effect of lowering the non-eternality rate represented in the left of Figure~\ref{fig:scatter_144_145}.
We cannot infer independence of $m_v$ without doing a statistical test; let our competing hypotheses be
(H0) for given $m_h$, all batches conditioned on $m_h$ with varying $m_v$ have the same rate, with deviations owing to chance; and
(H1) the rate parameter varies with $m_v$.
\noindent We test this hypothesis using likelihood-ratios, with the numerater the likelihood maximized over a single rate parameter for all $m_v$, and the denominator maximized over separate rate parameters for each batch.\footnote{If $s_i$ events are observed in a sample of size $n_i$, with $i$ indexing values of $m_v$, then we have for the log-likelihood ratio
	\begin{equation}
	\tau = 2 \log \frac{\ell_0}{\ell_1} \qquad \ell_h = \begin{cases}{\underset{p}{\mathrm{max}} \prod_i \mathrm{Beta}(p,s_i+1,n_i-s_i+1)}, & h = 0 \\ {\underset{\{p_i\}}{\mathrm{max}} \prod_i \mathrm{Beta}(p_i,s_i+1,n_i-s_i+1)}, & h = 1 \end{cases} \label{likerat} \end{equation}
	We then compute the distribution over likelihood ratios in samples with the same sample sizes as the original batches, given the single rate that maximized likelihood in the null hypothesis.}  We found that generally we cannot reject the null hypothesis with an alpha of 0.01.

Since we cannot conclude that changing the scale of the potential in this range has an effect on the rate of stochastic inflation, we also depict combined results taking models from the full range of $m_v$ (including mass bins with no positive events) as belonging to one sample.
\[ P(m \in D' \mid m \in S \cap \{m_h\} \cap \{n_s,\alpha\} ) \]
These are the shaded bars in the left plot of Figure~\ref{fig:scatter_144_145}.

\section{Matching Observables}
We are most concerned with the subpopulation of observationally viable inflation cosmologies. Here we depict how those models are distributed within our window onto model space (rectangular in the mass scales $m_v$ and $m_h$).
These results have only to do with ordinary inflation, so they are roughly commensurate those presented in \cite{tegmark08} for Measures A and B.
%

\begin{figure}[h!]
\centering
\includegraphics[height=5cm]{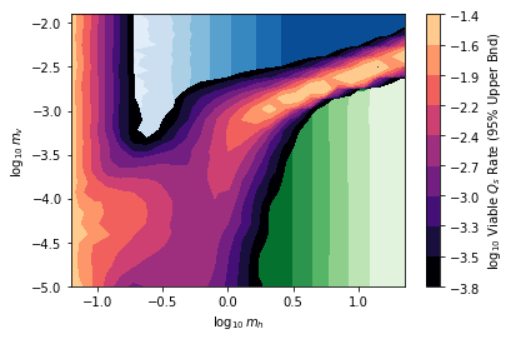} 
\includegraphics[height=5cm]{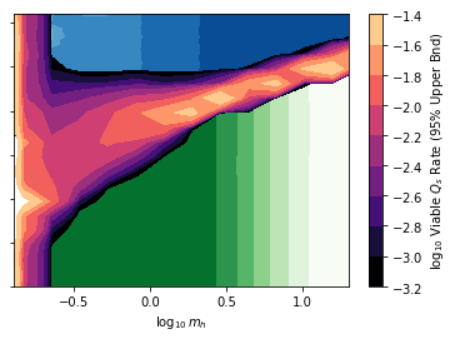} 

\includegraphics[width=6.5cm]{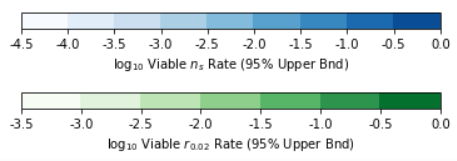} 
\caption[Marginal distributions of spectral parameters in Measures A and B.]{Marginal distributions of spectral parameters in Measures A (left) and B (right).  Foreground (magma): 95\% confidence upper bound on the rate of incidence of $Q_s$ falling within {\it Planck 2018} 68\% confidence interval.  The vertical striation pattern emerging on the left-hand side reflects the shortage of samples with successful inflation at low $m_{\text{h}}$ -- due to slower accrual of $e$-folds and the second slow roll criterion not being met at the peak -- resulting in a weaker bound.  Background: rates of incidence of $n_s$ (blue, upper) and $r$ (green, lower) falling within {\it Planck}'s 68\% confidence intervals, with higher color saturation (darker gray) indicating a higher rate.}
\label{fig:142_Q}
\end{figure}

\paragraph{Measure A}
Figure~\ref{fig:142_Q} depicts the 95\% confidence upper bounds on marginal rates of incidence for parameters describing the CMB power spectrum falling within {\it Planck 2018} 68\% confidence intervals, calculated by the procedure described in Appendix \ref{app:stats_methods}.
Attending to the super-Planckian regime $m_h \gtrsim 3$: note the positive correlation of mass scales along tightly spaced contours defining equal rates of incidence for matching of the scalar amplitude $Q_s$.
With $Q_s$ going like $m_v^2 m_{\text{h}} \times \mathcal O(10^{-2})$, one might expect that correlation to be negative---why the inversion?
Since we select for potentials with a small vacuum energy in the final basin and then shift $\rho_\Lambda$ to zero, inflation always ends.
At this scale inflation almost always ends very close to the minimum, where the potential is approximately quadratic and perturbation spectral parameters take their familiar forms for $V(\varphi) \sim \varphi^2$. 
In this regime, we more efficiently retain small fluctuations as $m_v$ increases by delaying the end of inflation---drawing the horizon exit scale closer to the minimum where 
\[ f(x)^{3/2} \abs{f'(x)}^{-1} \sim f(x) \]
is already very small---rather than reducing $m_{\text{h}}$ to make small values of $Q_s \sim m_v^2 m_{\text{h}}$ more likely far from the minimum where $V(\varphi) \sim m_v^4$.
Furthermore, larger field scales are more likely to yield large inflating intervals contiguous with the maximum---bridging multiple smaller disjoint intervals, and giving a slow roll streak starting from the peak access to lower intervals of $V(\varphi)$ where $Q_s$ can be small.

The constraints on $n_s$ are easily satisfied for large $m_{\text{h}}$ in Measure A, as they encompass the quadratic limit at 55 $e$-folds before inflation's end; but a sufficiently small tensor-to-scalar ratio is hard to come by in that regime.
\[ \text{Quadratic limit ($\mathcal N_e = 55$):} \qquad n_s \approx 1 - 2 \mathcal N_e^{-1} = 0.963 \qquad r \approx 8 \mathcal N_e^{-1} \approx 0.15 \]
%
At intermediate scales $0.1 \lesssim m_{\text{h}} \lesssim 1$, it is no longer guaranteed that inflation continues all the way from the maximum to the quadratic neighborhood of the minimum; peaks must be low enough that horizon exit occurring high on the potential can still produce small curvature perturbations.  
For field scales more than an order of magnitude {smaller} than $m_{\text{P}}$, we run into issues of sample size that limit our ability to assign a small upper bound on the rate estimate, 
reflecting the difficulty of finding potentials varying on sub-Planckian scales that produce enough $e$-folds of inflation.  This is acceptable for our purposes in Measure A, as only one in $\sim10^{4}$ successful models at that scale fall in the confidence region for the spectral index, and so models from smaller field scales are unlikely to significantly affect results close to home in the space of observables.



\paragraph{Measure B}
Measure B produces similar distributions to Measure A in the large field regime, as whether or not inflation starts at a maximum makes no difference if it persists along most of the potential slope to end close to the minimum.  
Approaching the Planck scale $m_h = 1$, we do not see a pronounced entry of models with smaller potential scales producing sufficiently small scalar perturbations as in Measure A.
Inflation is highly concentrated around extremal points of the potential, and with no guarantee of starting in one of those intervals, we do not get much inflation with horizon exit occurring in the intervening part of the slope, where small $m_v$ would give small $Q_s$.

Small sample size due to low rates of successful inflation becomes limiting at a larger field scale than in Measure A, since we are no longer initializing in a slow roll interval in every case.
Among the successful models, the scalar tilt is significantly more likely to fall in the observed range at small field scales, as most Measure B models in that regime feature an extended slow roll plateau rather than merely a gently curved quadratic peak.
(This is because the probability of sampling the initial field value within a slow roll interval is proportional to its width, and the number of $e$-fold counting toward the horizon problem threshold is not taken to be infinite as in Measure A.)
Meanwhile, the small tensor amplitude becomes \emph{slightly} harder to come by at large $m_{\text{h}}$, as there is some probability of large-field potentials eligible for inclusion in Measure A to be omitted from Measure B if the field value is sampled too close to the minimum.

\paragraph{Measure C}
Measure C departs from A only in cases for which fewer than 55 $e$-folds elapse beyond a 1-$\sigma$ deviation for Hubble-scale fluctuations in the direction of the Minkowski basin, so that horizon exit occurs in the excluded interval around the peak.
Since the size of those fluctuations is typically much smaller than the field scale in this range, in those models inflation is almost entirely localized at the maximum.
So we may expect a departure in the far end of the small-field regime, where inflation is localized at the peak and the scalar amplitude computed in that neighborhood can be sufficiently small (lower left corner of the lefthand plot in Figure~\ref{fig:142_Q}).


\end{document}